\documentclass[a4paper,11pt]{article}
\pdfoutput=1 

\usepackage{jcappub} 

\usepackage[T1]{fontenc} 

\usepackage{enumerate}
\usepackage{slashed}

\title{Particle Production by Time-Varying Dark Energy and the End of Cosmic Expansion}

\abstract{
    We consider various possible consequences of time-varying dark energy due to a quintessence scalar field whose energy density is partially converted to particles as the field evolves down its potential. This particle production acts as a source of thermal friction on the field that can make it difficult to distinguish whether dark energy is due to a radiating field rolling down a steep potential, a purely self-interacting field moving down a flatter potential, or a cosmological constant. By reducing the acceleration of the scalar field, thermal friction increases the amount of accelerated expansion and can cause a sizable bump in the quintessence equation of state. We take special interest in the case where a steep potential rapidly changes from positive to negative as the field evolves, resulting in the end of cosmic expansion and the beginning of contraction. Even in this case, we find that thermal friction lengthens the period of accelerated expansion and consequently delays the end of cosmic expansion, making it challenging to detect the impending transition to contraction using conventional cosmological tests. However, particle production can also provide alternative avenues for detection by generating a background of thermal dark radiation, partly comprised of neutrinos or other particles, whose energy density exceeds the remnant photon energy density.}


\author{Nicolas Patino}
\author{and Paul J. Steinhardt}
\affiliation{Department of Physics, Princeton University, Princeton, New Jersey 08544, USA}

\emailAdd{npatino@princeton.edu}
\emailAdd{steinh@princeton.edu}

\begin{document}

\maketitle
\flushbottom

\newcommand{\Physics}{\affiliation{Department of Physics, Princeton University, Princeton, New Jersey 08544, USA}}

\author{Nicolas Patino \orcidlink{0000-0002-3155-9750}}
\Physics

\author{Paul J. Steinhardt \orcidlink{0000-0003-3488-1603}}
\Physics


\section{Introduction} \label{sec:intro}
\allowdisplaybreaks[4]

The observed accelerated expansion of the universe is commonly attributed to some form of dark energy, but whether this accelerated expansion will continue eternally, or is even occurring at present \cite{N_equations,Shlivko_2025}, is an unsettled issue. Dark energy is often modeled as a cosmological constant, which if true would result in eternal accelerated expansion, but the most recent observational results from DESI \cite{DER} indicate dark energy may be time-varying, allowing other possible futures. Moreover, theoretical motivation for considering time-varying dark energy comes from two independent sources: quantum gravity and the homogeneity, isotropy, and flatness problems. The Trans-Planckian Censorship Conjecture (TCC) from the swampland program \cite{TCC} claims that quantum gravity imposes a limit on the net amount of accelerated expansion the universe can undergo and therefore forbids dark energy from being a cosmological constant. Similarly, cyclic bouncing models of the universe \cite{cyclic, IFS} posit that the present phase of cosmic expansion cannot persist indefinitely and must transition to a period of slow contraction, leading to a future bounce.

If dark energy is time-varying, there are two possible futures for the universe: either cosmic expansion is eternal, or cosmic expansion ends. For dark energy due to a \textit{simple quintessence scalar field} -- one that has potential $V(\varphi)$ and is purely self-interacting -- each scenario is possible depending on the potential $V(\varphi)$. Two such examples are the following:
\begin{enumerate}[I.]
    \item A steep exponential potential
    \begin{equation} \label{p1}
        V_\text{I}(\varphi)=V_1 e^{-\lambda_1\varphi}
    \end{equation} 
    with $V_1>0$ and $\lambda_1 \geq \sqrt{2}$. Since $V_1>0$, the scalar field energy density is always positive, so cosmic expansion is eternal. (Here and throughout this paper, we use reduced Planck units, where $M_\text{Pl}^{-2}=8\pi G=1$.)
    \item A steep potential where a finite portion is well-described by a difference of exponentials
     \begin{equation} \label{p2}
        V_\text{II}(\varphi)=V_1 e^{-\lambda_1\varphi}-V_2e^{\lambda_2\varphi}
    \end{equation} 
    with $V_1, V_2>0$, $\lambda_1 \geq \sqrt{2}$, and $\lambda_2 >\sqrt{6}$. The potential rapidly changes from positive to negative values as the field evolves, resulting in the end of cosmic expansion and the beginning of a phase of slow contraction.
\end{enumerate}
The TCC forbids shallow positive exponential potentials with $\frac{\left|V'\right|}{V}< \sqrt{2}$ \cite{TCC}. $V_\text{I}$ and $V_\text{II}$ are TCC-allowed potentials, and $V_\text{II}$ is consistent with cyclic bouncing models \cite{End}, but for $\lambda_1\geq \sqrt{2}$, both potentials violate observational constraints on the dark energy equation of state \cite{swampland, lambda_constraints, N_equations}, assuming the scalar field is purely self-interacting.

In this paper, however, we consider the consequences if the quintessence scalar field is coupled to light fields such that, as the scalar field evolves down its potential, its energy density is partly converted to relativistic particles that comprise a background of thermal dark radiation. The cosmological signatures of this form of dark energy have been explored in \cite{DER,cosmology_der,DER_Data} for a linear potential up to present day. Here, we investigate the effects of this particle production for potentials $V_\text{I}$ and $V_\text{II}$ up to and beyond present day, and in particular, we compare the results for each potential to determine whether an observable signature presently exists for the end of cosmic expansion. For $V_\text{II}$, we primarily limit our analysis to the \textit{expansion phase} and determine how particle production from time-varying dark energy might impact the evolution of the universe up to the end of cosmic expansion.

For this work, we consider two particle production mechanisms: a generic temperature-independent mechanism and an explicitly temperature-dependent mechanism, as proposed in \cite{DER}, where quintessence scalar field energy is converted to gauge bosons and other particles through sphaleron processes. In both cases, the particle production acts as a source of thermal friction on the scalar field, making it difficult to distinguish whether dark energy is due to a radiating field rolling down a steep potential, a purely self-interacting field rolling down a flatter potential, or a cosmological constant. For the temperature-dependent mechanism, we find that thermal friction can result in a sizable bump in the quintessence equation of state. More generally, particle production can significantly extend the duration of accelerated expansion and consequently delay the end of cosmic expansion, making it challenging to detect the impending transition using standard cosmological tests.

Though thermal friction can cause the expansion rate to resemble that of $\Lambda$CDM, the production of a background of dark radiation, which can contain neutrinos, provides alternative detection avenues. Due to increased particle production up to and during dark energy domination, the dark radiation density can grow to be much larger than the remnant photon energy density at present, as shown for a linear potential in \cite{DER}. The discovery of such a component could be an indirect indicator of the time-variation of dark energy \cite{cosmology_der} and the end of cosmic expansion.

The paper is organized as follows: In Section \ref{sec:2}, we review thawing quintessence with no particle production for potentials $V_\text{I}$ and $V_\text{II}$. In Section \ref{sec:3}, we describe the temperature-independent and temperature-dependent mechanisms by which the energy density of the quintessence scalar field can be converted to particles. Sections \ref{sec:acc_end} and \ref{sec:end} present the numerical analysis for the two kinds of potentials, respectively, when particle production of both the temperature-independent and temperature-dependent types are considered. Finally, in Section \ref{sec:discussion}, we conclude with a brief summary and a discussion of the implications for future theory and observations.  
\section{Simple Quintessence} \label{sec:2}
\allowdisplaybreaks[4]

We model time-varying dark energy as a homogeneous scalar field $\varphi$ moving down its potential $V(\varphi)$. In this section, we review the case of a simple quintessence scalar field, one that is minimally coupled, purely self-interacting, and described by action $S=\int\sqrt{-\text{det}(g_{\mu\nu})}\mathcal{L_\varphi}d^4x$, where the Lagrangian is 
\begin{equation}
    \mathcal{L_\varphi}=-\frac{1}{2}\partial_\mu\varphi\partial^\mu\varphi-V(\varphi),
\end{equation}
and $g_{\mu\nu}$ is the Friedmann-Robertson-Walker (FRW) metric in the $(-,+,+,+)$ signature. For a homogeneous scalar field, the equation of motion is then
\begin{equation}
    \ddot\varphi+3H\dot\varphi+V_{,\varphi}=0,
\end{equation}
where $H$ is the Hubble parameter, and $V_{,\varphi}$ denotes a derivative of the potential with respect to $\varphi$.

We assume a spatially flat universe with the observed quantities of matter and dark energy. The Friedmann equation is
\begin{equation}
    H^2=\frac{1}{3}\rho_\text{TOT}=\frac{1}{3}\left(\frac{\rho_{m,0}}{a^{3}}+\frac{1}{2}\dot\varphi^2+V(\varphi) \right),
\end{equation}
where $\rho_\text{TOT}$ and $\rho_\varphi= \frac{1}{2}\dot\varphi+V(\varphi)$ are the total and scalar field energy density, respectively; $\rho_{m,0}$ is the current (pressure-less) matter density; and $a$ is the scale factor with $a(t=t_0)=1$ corresponding to present day. We ignore the cosmic microwave background (CMB) and cosmic neutrino background (C$\nu$B) since we begin our numerical analysis well into matter domination where their impact on the numerical solutions is negligible. We consider the two types of potentials given by Eqs. (\ref{p1}) and (\ref{p2}): a steep exponential potential $V_\text{I}$ that is positive for all $\varphi$ and a difference of exponentials potential $V_\text{II}$ that runs from positive to negative.

For potential $V_\text{I}$, we start with ``thawing'' quintessence initial conditions, where the scalar field begins frozen by Hubble friction during matter domination and slowly gains kinetic energy over time. While the energy density of the scalar field is dominated by its potential, the equation of state of the scalar field 
\begin{equation}
    w_\varphi = \frac{\frac{1}{2}\dot\varphi^2-V(\varphi)}{\frac{1}{2}\dot\varphi^2+V(\varphi)}
\end{equation} 
is approximately $-1$. The second Friedmann equation is given by
\begin{equation}
    \ddot a=-\frac{a}{6}\left(\rho_\text{TOT}+3p_\text{TOT}\right)=-\frac{a\rho_\text{TOT}}{6}\left( 1+3w_\text{TOT}\right),
\end{equation}
where $p_\text{TOT}$ and $p_\varphi=\frac{1}{2}\dot\varphi^2-V(\varphi)$ are the total and scalar field pressure, respectively, and the total equation of state $w_\text{TOT}$ is defined as
\begin{equation}
    w_\text{TOT}\equiv\frac{p_\text{TOT}}{\rho_\text{TOT}}=\sum_iw_i\Omega_i,
\end{equation}
where the sum is over each component $i$ of the total energy density and $\Omega_i \equiv \frac{\rho_i}{3H^2}$. 

As the dark energy density grows to be comparable to the matter density, the total equation of state $w_\text{TOT}$ becomes less than $-\frac{1}{3}$, so by the second Friedmann equation, a phase of accelerated expansion begins. As dark energy begins to dominate and the matter density is diluted, the scalar field approaches its attractor solution, where $w_\text{TOT}= w_\varphi = \frac{1}{3}\lambda_1^2-1$. Since $\lambda_1 \geq \sqrt{2}$, accelerated expansion ends, as required by the TCC \cite{TCC}. Since the total energy density $\rho_\text{TOT}$ is always greater than 0, expansion continues eternally.

\begin{figure}[t]
    \centering
    \includegraphics[width=0.8\textwidth]{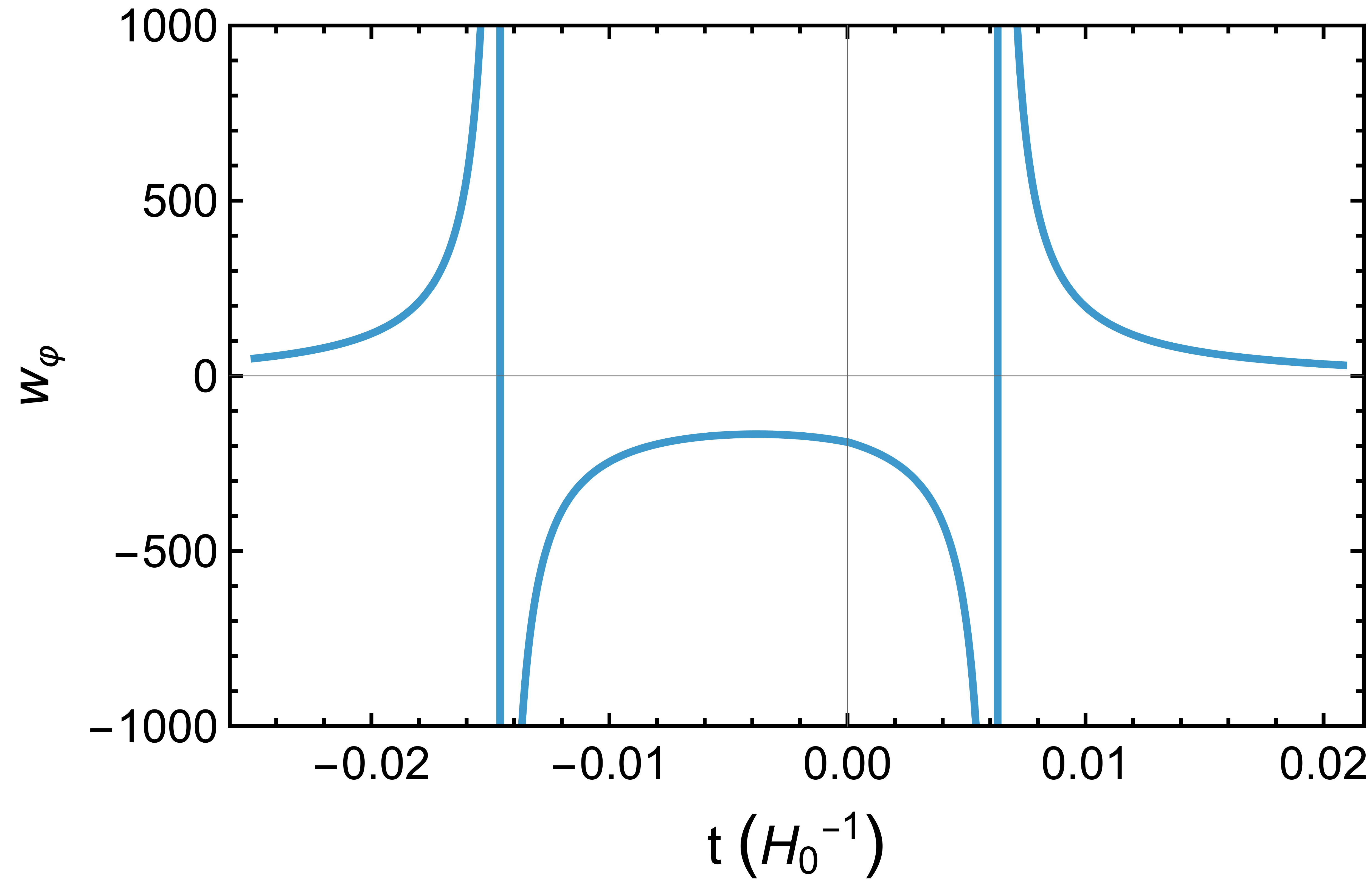}
    \caption{Quintessence equation of state $w_\varphi$ for a simple quintessence scalar field as a function of FRW time $t$, where $t=0$ corresponds to the end of cosmic expansion and the beginning of contraction. The two divergences show the two points when $\rho_\varphi=0$. During expansion, $\rho_\varphi$ transitions from positive to negative, and afterwards during contraction, $\rho_\varphi$ transition back from negative to positive. A similar sequence of divergences is shown in \cite{phantommatters} for the dark energy density of a scalar field coupled to matter.}
    \label{fig:1}
\end{figure}
For potential $V_\text{II}$, we again begin with thawing quintessence initial conditions, and the scalar field evolves in the same fashion. The main differences take place after accelerated expansion ends. As the field gains kinetic energy and moves from the positive region of the potential to the negative region, the expansion of the universe is decelerating. Due to Hubble friction, $\rho_\varphi$ transitions from positive to negative, causing the first divergence in the quintessence equation of state shown in Fig. \ref{fig:1}. If the potential is sufficiently negative, enough such that $\rho_\text{TOT}=0$, then cosmic expansion ends. At this point, $w_\text{TOT}$ diverges since $\rho_\text{TOT}=0$.

The universe then begins contracting. The total energy density $\rho_\text{TOT}$ increases monotonically from zero, so $\rho_\text{TOT}$ is never negative. Due to Hubble anti-friction, $\rho_\varphi$ increases from negative to positive, resulting in the second divergence shown in Fig. \ref{fig:1}. When $V_\text{II}(\varphi) \approx -V_2e^{\lambda_2\varphi}$, the scalar field approaches the attractor solution $w_\varphi=\frac{1}{3}\lambda_2^2-1$. Since $\lambda_2 > \sqrt{6}$, the scalar field energy density grows more quickly than spatial curvature and anisotropy. The universe is consequently smoothed and flattened by ``slow'' contraction.

The future of the universe then depends on the shape of the potential at larger values of $\varphi$ and other components of the theory. In many cyclic bouncing models \cite{cyclic, IFS}, the potential has a global minimum and eventually returns to positive values. Added contributions to the total action can then enable a bounce from slow contraction to expansion \cite{Bouncing}. Alternatively, the universe could contract to a crunch. In this paper, we primarily focus on the expansion phase in the case of $V_\text{II}$ and study how particle production from time-varying dark energy might impact the evolution of the universe and the end of cosmic expansion. 
\section{Particle Production Mechanisms} \label{sec:3} 
\allowdisplaybreaks[4]

\subsection{Particle Production Set-Up}
In this section, we describe how the equations of motion for a quintessence scalar field are affected if it has couplings to light fields that enable the partial conversion of scalar field energy density into relativistic particles that comprise a background of dark radiation, as explored in \cite{DER,cosmology_der}. We consider couplings such that the continuity equations for the scalar field and dark radiation are modified by the following dissipation term:
\begin{equation} \label{phi}
    \dot\rho_\varphi= -3H\dot\varphi^2-\Upsilon\dot\varphi^2
\end{equation}
and
\begin{equation} \label{DR}
    \dot\rho_\text{DR}= -4H\rho_\text{DR}+\Upsilon\dot\varphi^2,
\end{equation}
which can arise from microphysics such as sphaleron production due to an axion-like coupling \cite{DER}, as reviewed in Sec. \ref{sec:sub:TD}. Here, $\rho_\text{DR}$ is the dark radiation density, and $\Upsilon$ is the source of dissipation due to particle production. The form of $\Upsilon$ depends on the particle production mechanism. Eq. (\ref{phi}) can be equivalently expressed as the equation of motion for the scalar field:
\begin{equation} \label{eom}
    \ddot\varphi+3H\dot\varphi +\Upsilon\dot\varphi+V_{,\varphi}=0,
\end{equation}
where we see that the scalar field is now affected by Hubble friction from cosmic expansion (anti-friction during contraction), $3H\dot\varphi$, and thermal friction from particle production, $\Upsilon\dot\varphi$.

We assume that the background of dark radiation self-interacts sufficiently to be in thermal equilibrium, so the dark radiation density is given by
\begin{equation}
    \rho_\text{DR}=\frac{g_*\pi^2}{30}T_\text{DR}^4,
\end{equation}
where $T_\text{DR}$ and $g_*$ are the temperature and relativistic degrees of freedom of the dark radiation, respectively. We again ignore relic radiation from the CMB and C$\nu$B since their impact is negligible during the late matter-dominated and dark energy-dominated epochs ($z<200$). Assuming a flat universe with matter, the scalar field, and dark radiation, the Friedmann equation and matter continuity equation are then
\begin{equation}
    \begin{split}
     H^2 &=\frac{1}{3}(\rho_m+\rho_\varphi+\rho_\text{DR})\\ &=\frac{1}{3}\left(\rho_m+\frac{1}{2}\dot\varphi^2+V(\varphi)+\frac{g_*\pi^2}{30}T_\text{DR}^4\right)
    \end{split}
\end{equation}
and
\begin{equation}
    \dot\rho_m=-3H\rho_m,
\end{equation}
respectively.

To make these equations easier to solve numerically, we rewrite them in terms of $N$ instead of $t$, where
\begin{equation}
    N\equiv \ln(a)
\end{equation}
is the number of $e$-folds of expansion relative to present day; $a(t_0)=1$ corresponds to today ($t=t_0$), and we use $'$ to denote derivatives with respect to $N$, as done similarly in \cite{N_equations}:
\begin{equation} \label{eq:1}
    \varphi''+\frac{3}{2}\varphi'\left(1-w_\varphi\Omega_\varphi-\frac{1}{3}\Omega_\text{DR}\right)+\frac{\Upsilon}{H}\varphi'+\frac{V_{,\varphi}}{H^2}=0,
\end{equation}
\begin{equation} \label{eq:2}
    \Omega_\varphi'=3w_\varphi\Omega_\varphi^2+(\Omega_\text{DR}-3w_\varphi)\Omega_\varphi-\frac{\Upsilon\varphi'^2}{3H},
\end{equation}
\begin{equation} \label{eq:3}
    \Omega_\text{DR}'=\Omega_\text{DR}^2+(3w_\varphi\Omega_\varphi-1)\Omega_\text{DR}+\frac{\Upsilon\varphi'^2}{3H},
\end{equation}
and
\begin{equation} \label{eq:4}
    \Omega_m+\Omega_\varphi+\Omega_\text{DR}=1.
\end{equation}
By using the definition $\Omega_\varphi\equiv\frac{\rho_\varphi}{3H^2}= \frac{\frac{1}{2}\dot\varphi^2+V(\varphi)}{3H^2}$, we can express the Hubble parameter as
\begin{equation}
    H=\pm\sqrt{\frac{V(\varphi)}{3\Omega_\varphi-\frac{1}{2}\varphi'^2}}
\end{equation}
with $+$ during expansion and $-$ during contraction. Since $\Omega_\text{DR}\equiv \frac{\rho_\text{DR}}{3H^2}=\frac{g_*\pi^2}{90H^2}T_\text{DR}^4$, these equations can be numerically integrated to find solutions for $\varphi$, $\Omega_\varphi$, $T_\text{DR}$, and $\Omega_m$.

For a given $\Upsilon$ and initial $\Omega_\text{DR}$, the solutions to Eqs. (\ref{eq:1})$-$(\ref{eq:4}) are only sensitive to the particle content of the dark radiation through $g_*$, so any thermal dark radiation background with $g_*$ relativistic degrees of freedom gives the same cosmological solutions. Further, if $\Upsilon$ is temperature-independent, the solutions for $\varphi$, $\Omega_\varphi$, $\Omega_\text{DR}$, and $\Omega_m$ are independent of $g_*$, and only $T_\text{DR}$ depends on $g_*$. For the sake of concreteness, we consider the interesting example of a scalar field coupled to a non-Abelian gauge group that is beyond the Standard Model in a dark sector, described by the following terms in the Lagrangian \cite{DER}:
\begin{equation} \label{eq:Lex}
    \begin{split}
    \mathcal{L} \supset & -\frac{1}{2}\partial_\mu\varphi\partial^\mu\varphi - V(\varphi)-\frac{1}{2g^2}\text{Tr }G_{\mu \nu}G^{\mu \nu} \\
    & -\bar{\psi}(\slashed{D}+m)\psi-F(\varphi,A_\mu^a,\psi),
    \end{split}
\end{equation}
where $A_\mu^a$ are non-Abelian gauge fields with group $SU(n_c)$; $g$ and $G_{\mu \nu}$ are the coupling constant and field strength tensor, respectively; $\psi\equiv \psi_n$ is the collection of $n_f$ fermions with mass matrix $m$ and masses $m_n$, and $F$ is the coupling between the scalar field, the gauge fields, and the fermions. We assume that $\alpha\equiv \frac{g^2}{4\pi}\ll 1$ and $\alpha^2T_\text{DR}\lesssim  m_n \ll T_\text{DR}$ so that the gauge bosons are deconfined, the fermions remain relativistic, and the chiral charge built up by sphaleron processes is sufficiently depleted \cite{cosmology_der}.

We explore the possibility that this dark sector has couplings to Standard Model particles. In particular, we focus on the case where neutrinos are produced, as explored in \cite{cosmology_der}. For example, the fermions may be coupled to a right-handed neutrino $\nu_R$, which itself is coupled to the lightest Standard Model neutrino through a Dirac mass-mixing term \cite{cosmology_der}:
\begin{equation} \label{eq:L}
    \mathcal{L} \supset -\frac{1}{f_{\nu_R}}G^a_{\mu \nu}\psi^a \sigma_{\mu\nu}\nu_R-y h \bar{\nu}_L \nu_R-\frac{1}{2}m\bar\nu_R(\nu_R)^c+h\text{.}c.,
\end{equation}
where $y$ is the Yukawa coupling, $h$ is the Higgs field, and we assume both the lightest left-handed and right-handed neutrinos have a mass much less than an meV. For $n_c=2$ and $n_f = 3$, the dark radiation background is comprised of three gauge bosons, three fermions, three anti-fermions, a sterile neutrino, and the lightest Standard Model neutrino, each with two polarizations. Further, thermalized particles of the scalar field, produced by the inverse of the processes that convert scalar field energy into dark radiation, give another degree of freedom, which totals to $g_* = 21$ \cite{cosmology_der}.

\subsection{Temperature-Independent Particle Production Mechanism $(\Upsilon=\Upsilon_\beta)$}
Rather than specify the exact coupling between the scalar field and the relativistic degrees of freedom, we first consider a particle production mechanism where the thermal friction coefficient $\Upsilon$ is a positive constant, as in \cite{cosmology_der, DER_Data}. We study this scenario as a simple example to compare against an explicitly temperature-dependent mechanism that emerges from first principles \cite{DER}. Further, a constant $\Upsilon$ may be a good approximation for particle production mechanisms that are fairly insensitive to the dark radiation temperature, such as those with $\Upsilon\propto T_{\text{DR}}^\epsilon$ for $|\epsilon|\ll1$.

If $\Upsilon$ is much greater or much less than $H$, the resulting expansion rate is nearly indistinguishable from that of a cosmological constant or simple quintessence, respectively, up to present day. We take interest in cases where the thermal friction and Hubble friction are both non-negligible as the universe evolves. Thus, for this particle production mechanism, we choose
\begin{equation} \label{eq:TI}
    \Upsilon=\Upsilon_\beta \equiv \beta^2 H_0,
\end{equation}
where $\beta$ is a dimensionless constant.

\subsection{Temperature-Dependent Particle Production Mechanism $(\Upsilon=\Upsilon_\gamma)$} \label{sec:sub:TD}
As an alternative example, we consider a particle production mechanism where the friction coefficient $\Upsilon$ is explicitly temperature-dependent, as is the case when the scalar field has the following axion-like coupling to non-Abelian gauge fields \cite{DER,cosmology_der,warm_inflation,Hubble_tension, G19, J22}:
\begin{equation}
    \begin{split}
    F(\varphi,A^a_\mu,\psi) = \frac{\varphi}{f}\frac{\text{Tr }G^{\mu\nu}\Tilde{G}_{\mu\nu}}{16\pi^2},
    \end{split}
\end{equation}
where $f$ is the symmetry breaking scale. This coupling is compelling because the approximate shift symmetry of the scalar field suppresses thermal corrections to the potential \cite{cosmology_der}. In this scenario, the potential $V(\varphi)$ given in Eq. (\ref{eq:Lex}) is not the periodic axion potential that arises from instanton processes, but rather, it is a generic UV potential, as might arise from a soft symmetry breaking \cite{relaxion}, that has no restriction to be periodic or nonnegative. $V(\varphi)$ is the dark energy potential responsible for the observed accelerated expansion. We restrict $\alpha \ll 0.1$ in the temperature regime of the dark radiation. Due to the exponential suppression of the instanton scale, the axion periodic potential can be neglected relative to the dark energy potential $V(\varphi)$.

The equation of motion for the scalar field derived from this Lagrangian is then 
\begin{equation}
    \ddot \varphi+3H\dot\varphi+\frac{\text{Tr }G^{\mu\nu}\Tilde{G}_{\mu\nu}}{16\pi^2f}+V_{,\varphi}=0.
\end{equation}
If a thermal average is taken of this equation, sphaleron processes contribute a term proportional to $\dot\varphi$, which by the fluctuation-dissipation theorem \cite{FDT1, FDT2}, is:
\begin{equation} \label{eq:dissipation}
    \left\langle\frac{\text{Tr }G^{\mu\nu}\Tilde{G}_{\mu\nu}}{16\pi^2f} \right\rangle=\frac{\Gamma_\text{sph}}{2T_\text{DR}f^2}\dot\varphi,
\end{equation}
where the sphaleron rate is of order $\Gamma_\text{sph} \sim n_c^5\alpha^5T_\text{DR}^4$ up to logs of $\alpha$ \cite{sphalerons}. Sphaleron processes are efficient in the same regime when instanton processes are negligible: when $\alpha \ll 1$. As long as the temperature of the dark radiation is sufficiently larger than the deconfinement temperature of the gauge bosons $\alpha(T_c) = 1$, the equations above are valid \cite{cosmology_der,sphalerons}. Then, the friction coefficient has a temperature dependence
\begin{equation} \label{eq:TD}
    \Upsilon =\Upsilon_\gamma\equiv\gamma^2T_\text{DR}^3,
\end{equation}
where we define $\gamma\equiv \frac{1}{f}\sqrt{\frac{n_c^5\alpha^5}{2}}$.
\section{Eternal Cosmic Expansion $\left(V=V_\textmd{I}\right)$} 
\label{sec:acc_end}
\allowdisplaybreaks[4]

We first consider $V_\text{I}(\varphi)=V_1 e^{-\lambda_1\varphi}$ with $\lambda_1\geq \sqrt{2}$, a positive-definite potential that results in eternal cosmic expansion. For each particle production mechanism, we numerically solve the cosmological equations of motion: Eqs. (\ref{eq:1})$-$(\ref{eq:4}). We begin the numerical integration at $N=N_i$ well into matter domination when $\Omega_\varphi(N_i)=10^{-6}$. We approximate $\varphi'(N_i)=0$ since the scalar field is frozen by Hubble friction. The dark radiation density at this time is much less than the remnant photon density, so for the sake of illustration, we set $\Omega_\text{DR}(N_i)=10^{-8}$. We find that the solutions are largely insensitive to the value chosen for $\Omega_\text{DR}(N_i)$. We define present day ($N=0$) as the earliest time (or smallest $N$) where $\Omega_m(N)=0.3$. Finally, we use the shooting method to set $\varphi(N_i)$ such that $H(0)=H_0$, the present-day value of the Hubble parameter. The amount of particle production impacts what initial values $N_i$ and $\varphi(N_i)$ are needed to get $H(0)=H_0$ when $\Omega_m(0)=0.3$, so $N_i$ and $\varphi(N_i)$ vary depending on the value of $\beta$ or $\gamma$. For the plots in this section, we set $V_1 = 1$ $H_0^2M_{\text{Pl}}^2$ and $\lambda_1 = 1.5$.

\subsection{Temperature-Independent Particle Production Mechanism $\left(\Upsilon=\Upsilon_\beta\right)$}

\begin{figure}[t]
    \centering
    \begin{minipage}{.5\textwidth}
     \centering
     \includegraphics[width=1\linewidth]{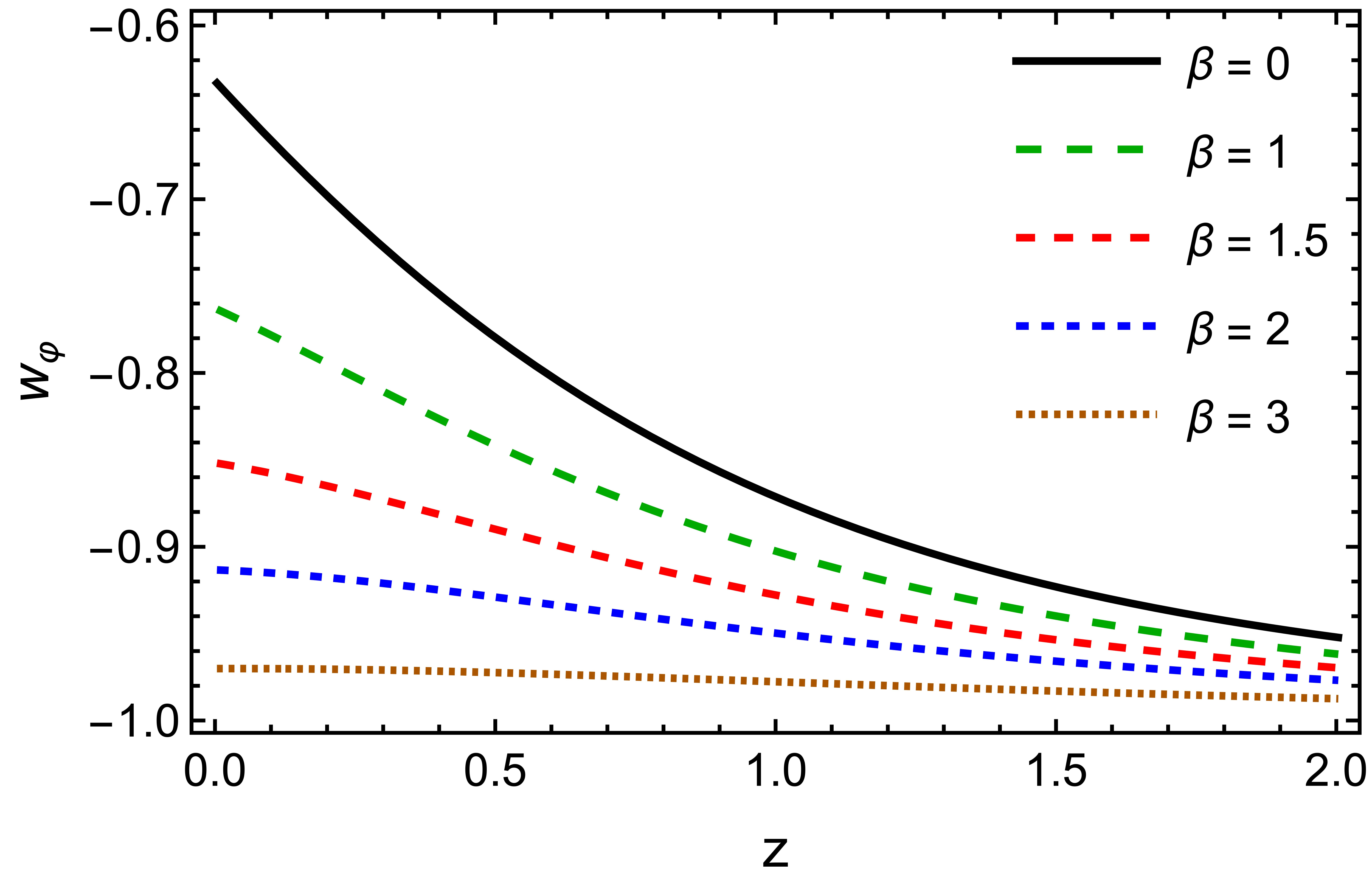}
    \end{minipage}%
    \begin{minipage}{.5\textwidth}
     \centering
     \includegraphics[width=1\linewidth]{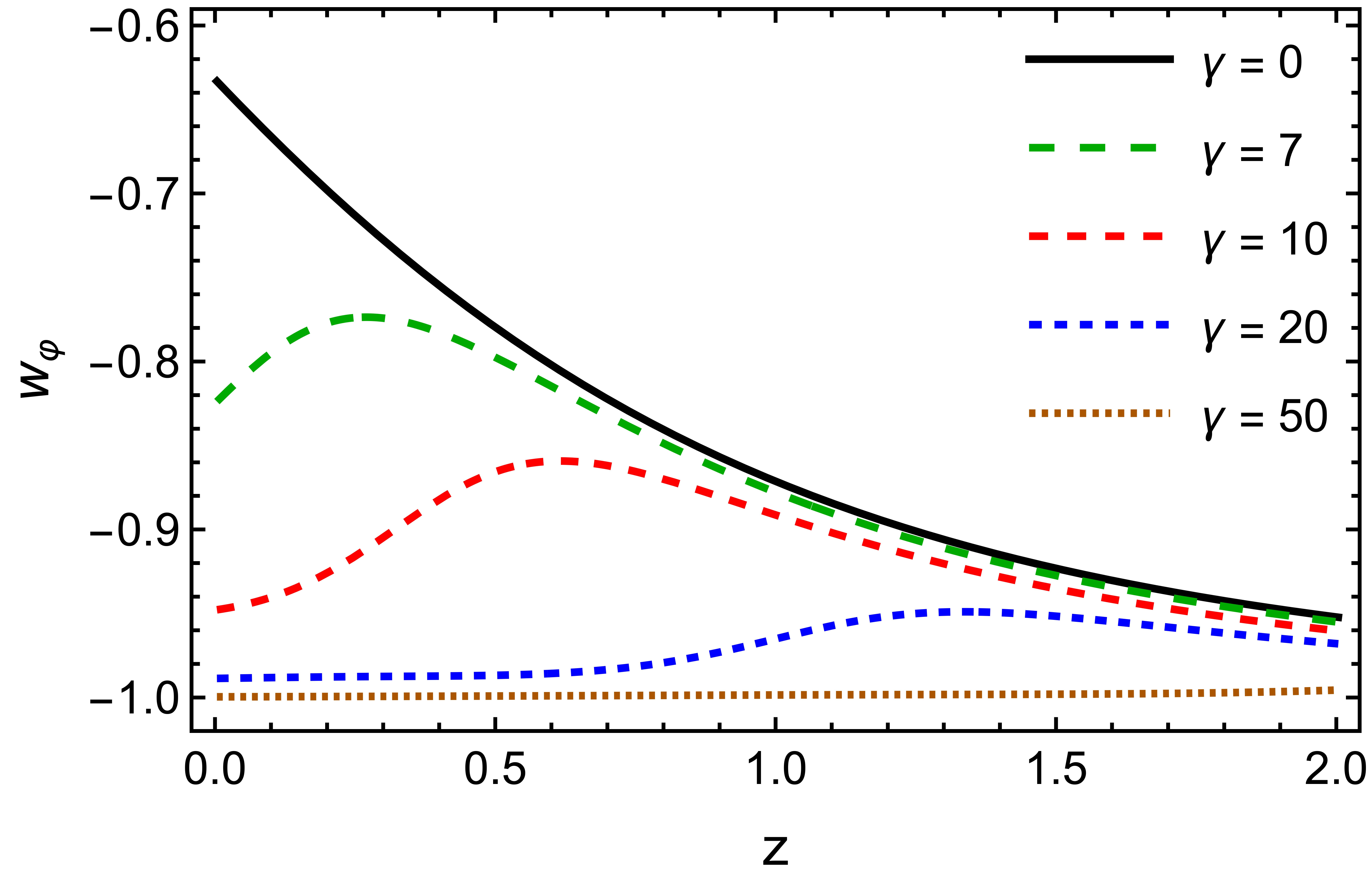}
    \end{minipage}
    \caption{Quintessence equation of state $w_\varphi$ as a function of redshift $z$ with exponential potential $V_\text{I}$ for the temperature-independent (left panel) and temperature-dependent (right panel) particle production mechanisms. Both plots compare the case of simple quintessence ($\beta =\gamma =0$) against cases where $\beta$ (unitless) and $\gamma$ (in units of $H_0^{-1/4}M_{\text{Pl}}^{-3/4}$) are non-zero. As seen in the right panel, there can be a sizable bump in the quintessence equation of state when the combined Hubble and thermal friction sufficiently decelerate the scalar field, which only occurs before present day for the temperature-dependent mechanism.}
    \label{fig:2}
\end{figure}

In the absence of particle production ($\beta =0$), the potential $V_\text{I}$ is too steep to produce sufficient accelerated expansion to be consistent with observational constraints on dark energy \cite{lambda_constraints,DESI}. The scalar field acceleration $\ddot\varphi$ is too large, resulting in $w_\varphi$ increasing above observational bounds. This issue, however, is alleviated by thermal friction from particle production, which reduces the acceleration of the scalar field. For larger $\beta$, thermal friction is greater and further reduces the rate at which $w_\varphi$ grows, as shown in the left panel of Fig. \ref{fig:2}.
\begin{figure}[t]
    \centering
    \begin{minipage}{.5\textwidth}
     \centering
     \includegraphics[width=1\linewidth]{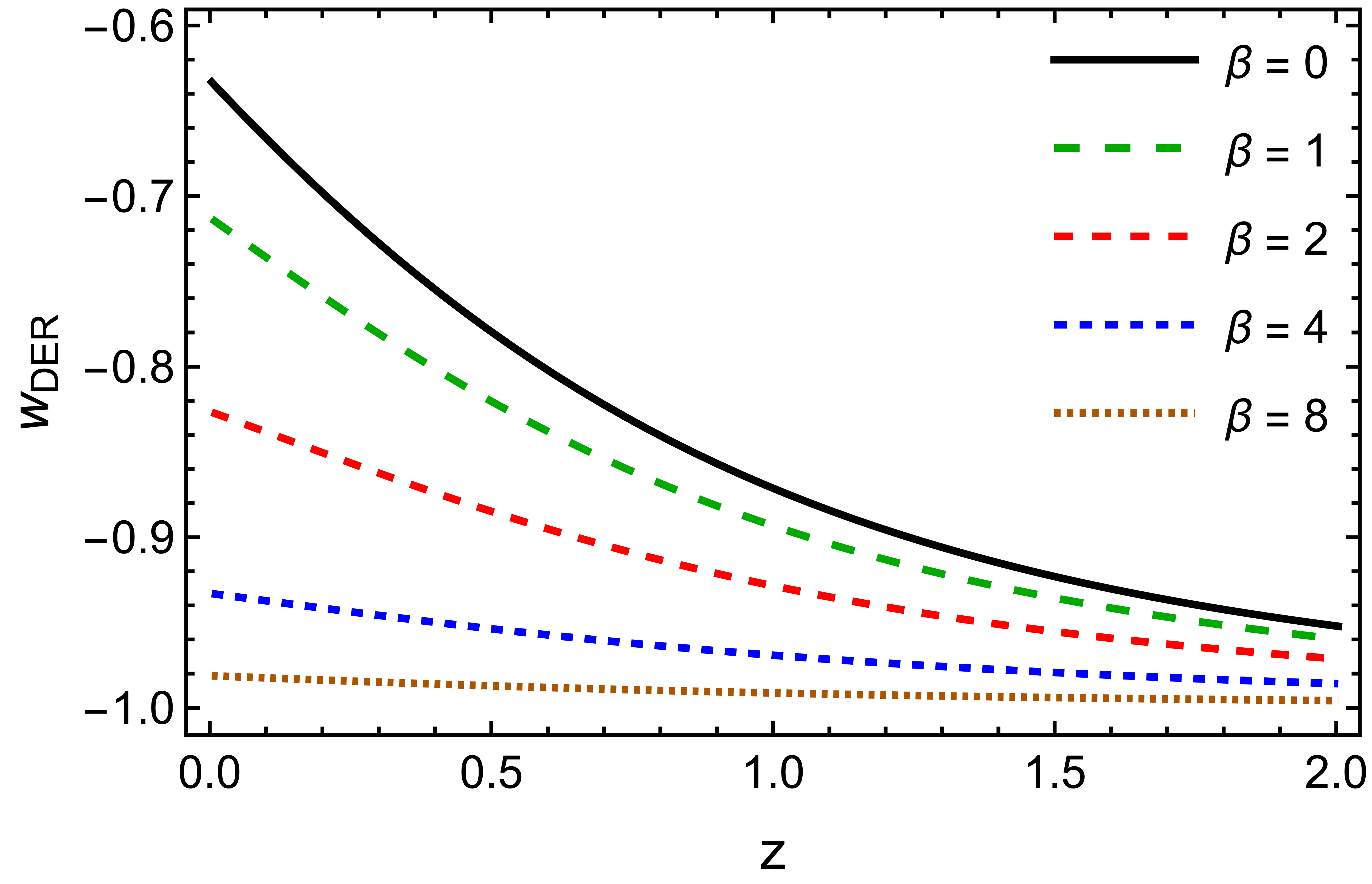}
    \end{minipage}%
    \begin{minipage}{.5\textwidth}
    \centering
    \includegraphics[width=1\linewidth]{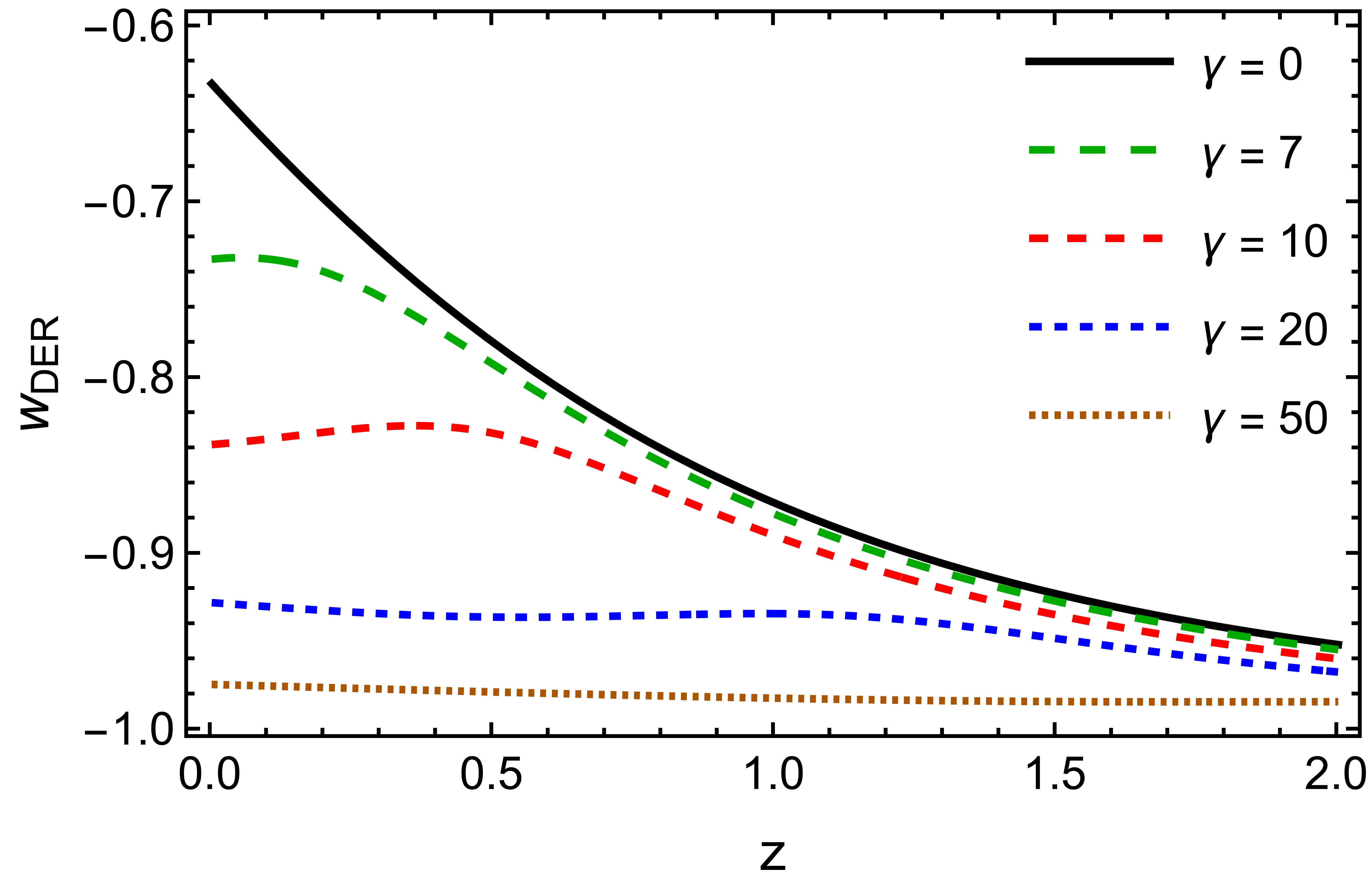}
    \end{minipage}
    \caption{Combined quintessence and dark radiation equation of state $w_\text{DER}$ as a function of redshift $z$ with exponential potential $V_\text{I}$ for the temperature-independent (left panel) and temperature-dependent (right panel) particle production mechanisms. Both plots compare the case of simple quintessence ($\beta =\gamma =0$) against cases where $\beta$ (unitless) and $\gamma$ (in units of $H_0^{-1/4}M_{\text{Pl}}^{-3/4}$) are non-zero. For both mechanisms, $w_\text{DER}$ can resemble the equation of state for a cosmological constant $w_\Lambda =-1$ if $\beta$ or $\gamma$ is large enough.}
    \label{fig:3}
\end{figure}
Even when accounting for the background of dark radiation, the left panel of Fig. \ref{fig:3} illustrates how the combined equation of state of the scalar field and dark radiation \cite{DER}
\begin{equation}
    w_\text{DER}=\frac{p_\varphi+p_\text{DR}}{\rho_\varphi+\rho_\text{DR}}= \frac{\frac{1}{2}\dot\varphi^2-V(\varphi)+\frac{1}{3}\rho_\text{DR}}{\frac{1}{2}\dot\varphi^2+V(\varphi)+\rho_\text{DR}}
\end{equation}
increases more slowly for larger $\beta$. \textit{For sufficiently large $\beta$, we find that steep TCC-allowed potentials, such as $V_\text{I}$, can have accelerated expansion similar to what occurs for flatter potentials or a cosmological constant.}

\begin{figure}
    \centering
    \begin{minipage}{.5\textwidth}
      \centering
      \includegraphics[width=1\linewidth]{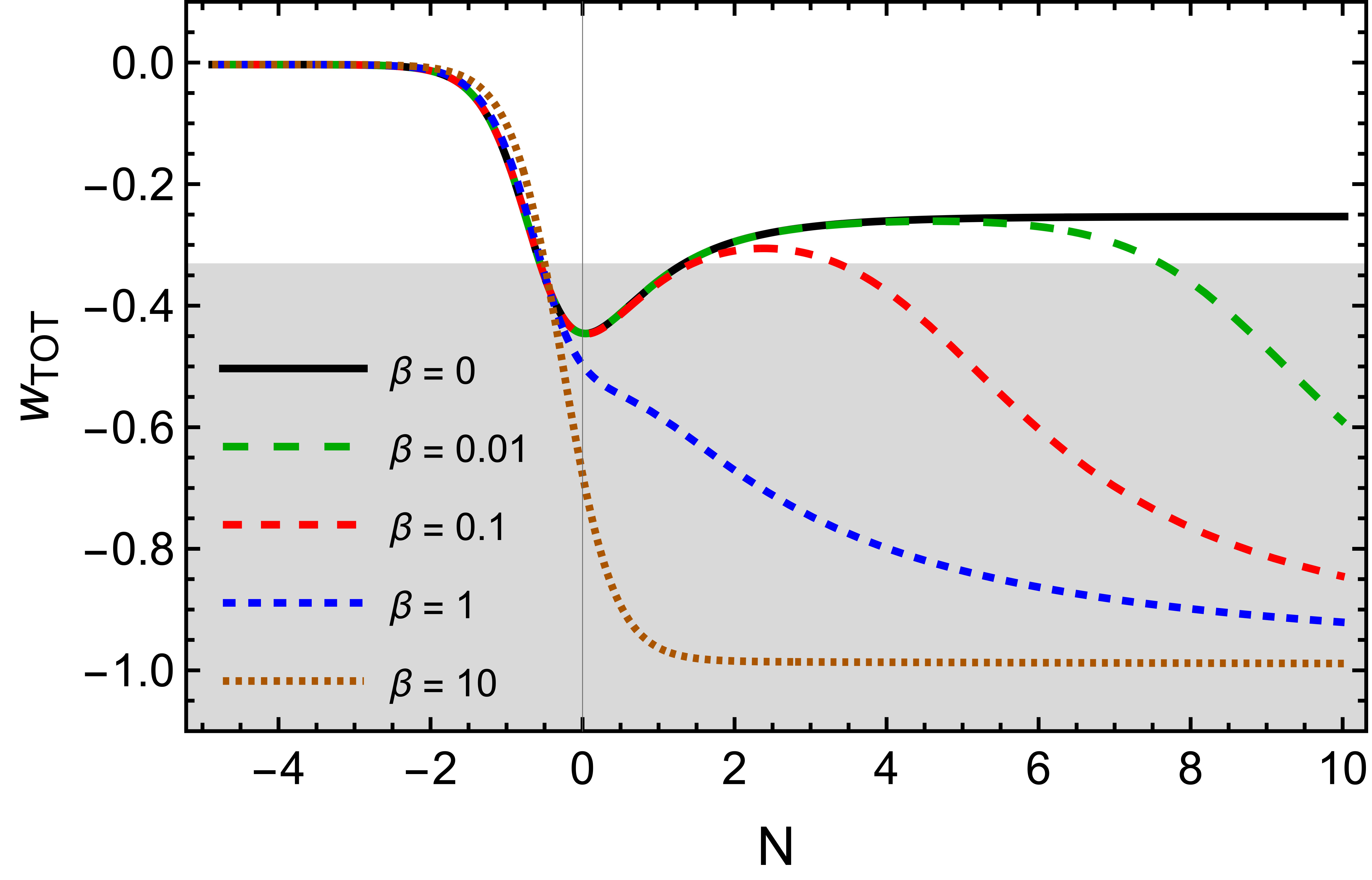}
    \end{minipage}%
    \begin{minipage}{.5\textwidth}
     \centering
     \includegraphics[width=1\linewidth]{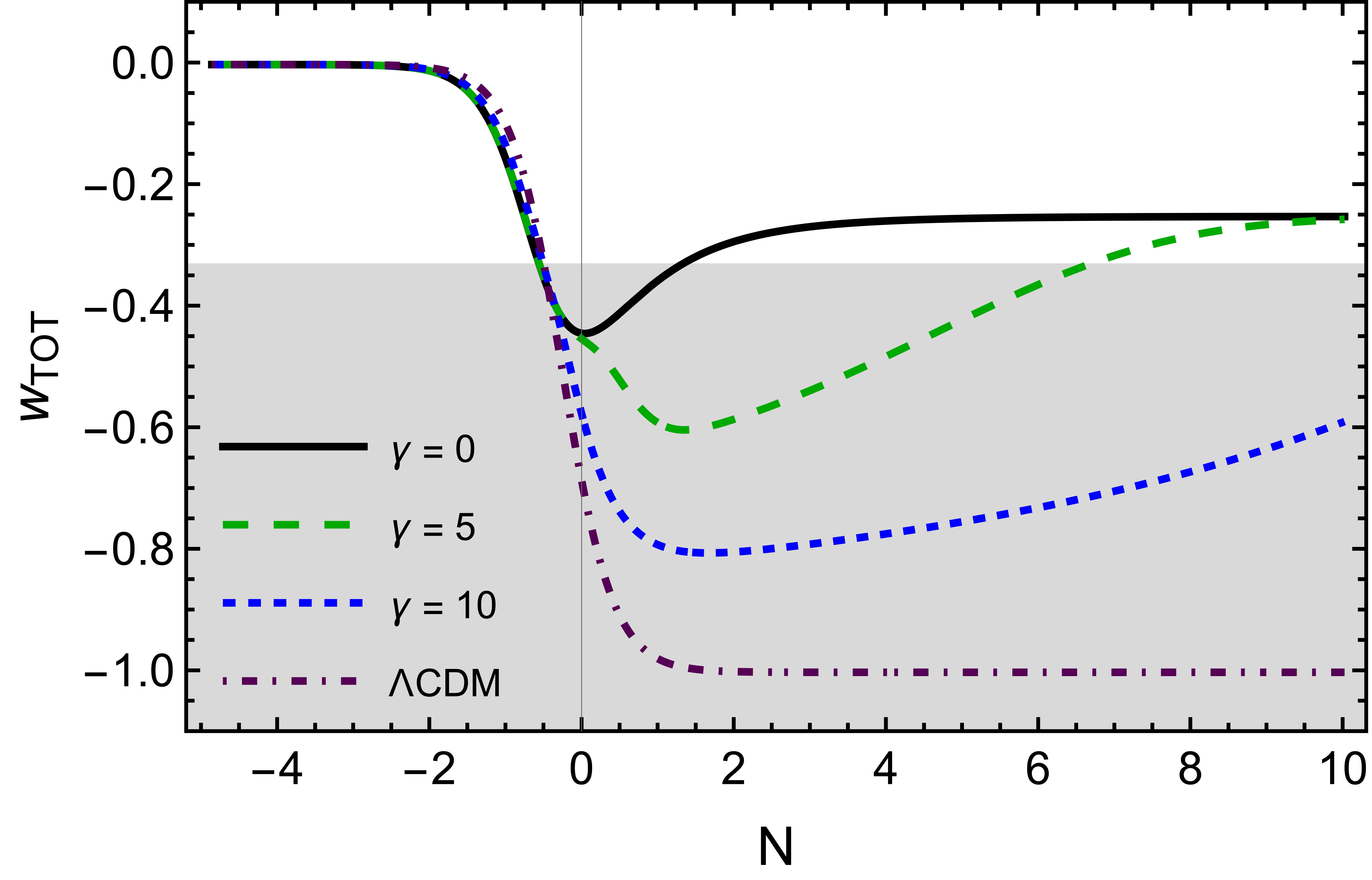}
    \end{minipage}
    \caption{Total equation of state $w_\text{TOT}$ as a function of $N$ $e$-folds relative to present day ($N=0$) with exponential potential $V_\text{I}$ for the temperature-independent (left panel) and temperature-dependent (right panel) particle production mechanisms. Both plots compare the case of simple quintessence ($\beta=\gamma=0$) against cases where $\beta$ (unitless) and $\gamma$ (in units of $H_0^{-1/4}M_{\text{Pl}}^{-3/4}$) are non-zero, and the right panel includes a curve for $\Lambda$CDM. When a curve is in the shaded region $\left(w_\text{TOT} < -\frac{1}{3}\right)$, the universe is undergoing accelerated expansion. For the temperature-independent mechanism (left panel), the present phase of accelerated expansion may end for a finite period in some cases, as shown for $\beta=0.01$ and $\beta=0.1$, but for all non-zero $\beta$, $w_\text{TOT}$ approaches $-1$ once thermal friction becomes significant relative to Hubble friction, resulting in eternal accelerated expansion. For the temperature-dependent mechanism (right panel), thermal friction increases the amount of accelerated expansion, but accelerated expansion always ends.}
    \label{fig:4}
\end{figure}

Beyond present day, temperature-independent particle production leads to a highly different future for the universe than simple quintessence. As seen in the left panel of Fig. \ref{fig:4}, simple quintessence ($\beta=0$) can lead to a brief period of accelerated expansion before the scalar field nears its attractor solution of $w_\text{TOT}=w_\varphi= \frac{1}{3}\lambda_1^2 -1$. The total equation of state approaches a constant value greater than $-1$ because the Hubble parameter, and consequently the Hubble friction, decreases as the scalar field energy density decreases. For $\lambda_1 > \sqrt{2}$, the universe transitions to an eternal period of decelerated expansion. 

For temperature-independent particle production, $w_\text{TOT}$ approaches $\frac{1}{3}\lambda_1^2-1$ during scalar field domination until the thermal friction is non-negligible relative to the Hubble friction. 
For smaller $\beta$, the Hubble friction dominates for a longer period. If $\beta$ is sufficiently small, the present accelerated expansion ends, leading to a temporary phase of decelerated expansion. Such cases are shown for $\beta =0.01 $ and $\beta =0.1$ in the left panel of Fig. \ref{fig:4}.

This decelerated expansion cannot continue eternally. Since $V_\text{I}$ is positive definite, the Hubble parameter will asymptote to zero while $\Upsilon_\beta$ remains constant. Eventually, thermal friction becomes significant relative to Hubble friction. Then, the scalar field no longer approaches its simple quintessence attractor solution. For constant $\Upsilon$, $\ln{\left(\frac{1}{2}\dot\varphi^2\right)}$ decreases more rapidly than $\ln{\left(V_\text{I}\right)}$, so the scalar field energy density becomes potential dominated and approaches $w_\text{TOT}=w_\varphi=-1$. For large enough $\beta$, present accelerated expansion never ends. Even for sufficiently small $\beta$, the temporary phase of decelerated expansion ends, and a new era of accelerated expansion begins. As in $\Lambda$CDM, the ultimate future of the universe is eternal accelerated expansion, regardless of the value of $\beta$. 

\begin{figure}[t]
    \centering
    \begin{minipage}{.5\textwidth}
      \centering
      \includegraphics[width=1\linewidth]{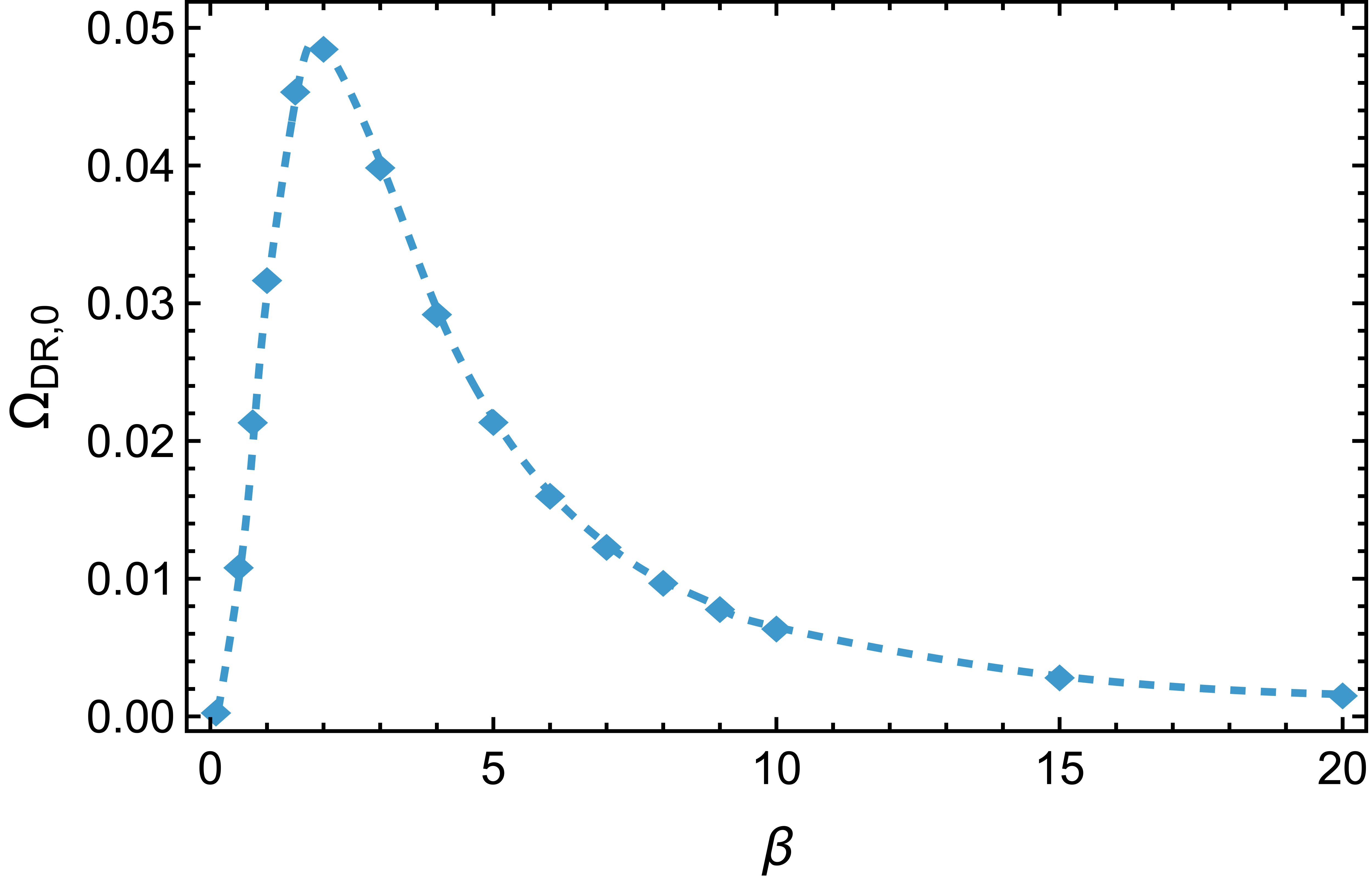}
    \end{minipage}%
    \begin{minipage}{.5\textwidth}
      \centering
      \includegraphics[width=1\linewidth]{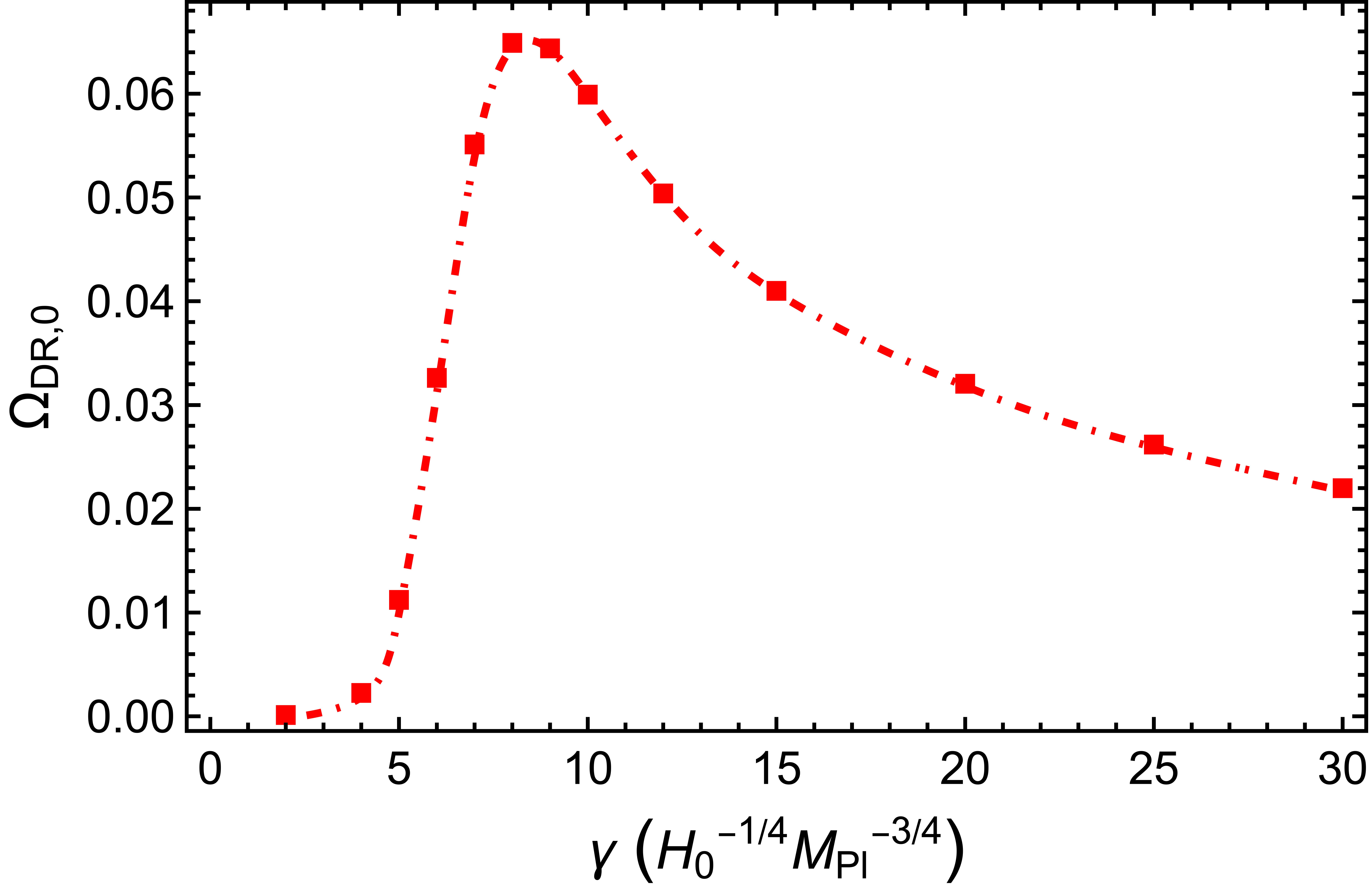}
    \end{minipage}
    \caption{Present-day dark radiation fractional density $\Omega_\text{DR,0}$ with exponential potential $V_\text{I}$ for the temperature-independent (left panel) and temperature-dependent (right panel) particle production mechanisms. The rate of dark radiation production $\Upsilon\dot\varphi^2$ is affected by $\beta$ and $\gamma$ through $\Upsilon$ directly and $\dot\varphi$ indirectly via thermal friction, so $\Omega_\text{DR,0}$ does not increase monotonically with $\beta$ and $\gamma$.}
    \label{fig:5}
\end{figure}

In many cases, we find that thermal friction from particle production can make time-varying dark energy difficult to distinguish from a cosmological constant using conventional cosmological tests. However, the production of a background of dark radiation provides an alternative detectable effect \cite{cosmology_der}. Depending on the value of $\beta$, the current dark radiation density can be much larger than the remnant photon energy density $\left(\Omega_{\gamma, 0}\approx 5 \times 10^{-5}\right)$, as illustrated by the left panel of Fig. \ref{fig:5}. Since most of the dark radiation is produced closer to present day, its density is not subject to the usual CMB constraints on effective relativistic degrees of freedom. For example, assuming a linear potential, cosmological data allow for, and can even prefer, a present dark radiation density up to $\Omega_\text{DR,0}=0.03$ \cite{cosmology_der, DER_Data}. If the lightest left-handed neutrino is part of the dark radiation, there could exist a background of this neutrino at a temperature on the order of a meV, larger than the temperature of the relic C$\nu$B. At these temperatures, this neutrino background can potentially be detected by experiments such as PTOLEMY \cite{ptolemy}, as explored in \cite{cosmology_der}. Such a discovery could be a sign of dynamical dark energy.

Contrary to what might be expected, the present-day dark radiation density does not increase monotonically with $\beta$, as shown in the left panel of Fig. \ref{fig:5}, because the production rate $\beta^2H_0\dot\varphi^2$ has a non-trivial dependence on $\beta$. Since $\beta$ impacts the acceleration of the scalar field through thermal friction, $\beta$ affects the dark radiation production rate through $\beta^2H_0$ directly and $\dot\varphi$ indirectly by affecting the dynamics of the scalar field. If $\beta$ is too large, the thermal friction reduces the acceleration of the field such that $\dot\varphi$ is small even by present day, resulting in less dark radiation production. If $\beta$ is too small, the scalar field is nearly unaffected by thermal friction, allowing $\dot\varphi$ to be large, but the rate of dark radiation production is still small because it is proportional to $\beta^2$. Thus, the present-day dark radiation density is maximized by a finite value of $\beta$.

\subsection{Temperature-Dependent Particle Production Mechanism $\left(\Upsilon=\Upsilon_\gamma\right)$}

For temperature-dependent particle production, we again find that, given a sufficiently large $\gamma$, $V_\text{I}$ can still result in accelerated expansion that resembles $\Lambda$CDM predictions. As seen in the right panels of Figs. \ref{fig:2} and \ref{fig:3}, thermal friction reduces the scalar field acceleration, resulting in a $w_\varphi$ and $w_\text{DER}$ that grow more slowly. Despite these similarities, we find that the temperature dependence of $\Upsilon_\gamma$ can lead to significant differences in how the scalar field equation of state $w_\varphi$ evolves over time. 

In the absence of particle production $(\gamma=0)$, a thawing quintessence scalar field initially has positive acceleration $\ddot \varphi > 0$ since $-V_{,\varphi}>3H\dot\varphi$. For a positive potential, $w_\varphi = \frac{\frac{1}{2}\dot\varphi^2- V(\varphi)}{\frac{1}{2}\dot\varphi^2+ V(\varphi)}=\frac{\frac{1}{2V(\varphi)}\dot\varphi^2- 1}{\frac{1}{2V(\varphi)}\dot\varphi^2+ 1}$ increases monotonically toward the attractor solution as $\dot\varphi$ increases and $V$ decreases. The addition of thermal friction does not necessarily change this sequence. For the temperature-independent mechanism, the combined Hubble and thermal friction do not sufficiently decelerate the field to significantly decrease $w_\varphi$ before present day, at least for the examples we considered. As seen in the left panel of Fig. \ref{fig:2}, $w_\varphi$ increases from large redshift up to present day. 

For the temperature-dependent mechanism, however, we find that $\Upsilon_\gamma\dot\varphi$ can grow quickly enough that the combined friction effects sufficiently decelerate the field, causing $w_\varphi$ to decrease. As particle production raises the temperature of the dark radiation, $\Upsilon_\gamma=\gamma^2T_\text{DR}^3$ grows as well, which can increase the dark radiation production rate $\Upsilon_\gamma\dot\varphi^2$ and further raise the temperature. As a result, the thermal friction term can grow rapidly enough such that the combined Hubble friction and thermal friction briefly decelerate the field, causing $\dot\varphi$ to decrease. If the deceleration is large enough, $w_\varphi$ begins to decrease. This dip in $w_\varphi(z)$ gives the impression of a bump, as seen for $\gamma = 7,10,$ and $20$ $H_0^{-1/4} M_\text{Pl}^{-3/4}$ in the right panel of Fig. \ref{fig:2}. 

After this point, the thermal friction is the dominant friction term in the equation of motion, which approximately cancels the potential term and causes the field to have small acceleration $\ddot \varphi$. As a result, $w_\varphi$ does not change much over time, producing the plateau region seen when $z<0.6$ for $\gamma = 20$ $H_0^{-1/4} M_\text{Pl}^{-3/4}$. If $\gamma$ is too large, such as with $\gamma = 50$ $H_0^{-1/4} M_\text{Pl}^{-3/4}$, $\dot\varphi$ is already so small that the bump in the quintessence equation of state is imperceptible. Though $w_\varphi(z)$ and $w_\text{DER}(z)$ can resemble $w_\Lambda=-1$ for large enough $\gamma$, perturbations of the scalar field may leave a visible imprint on the CMB for sufficiently large thermal friction \cite{cosmology_der}.

Unlike for temperature-independent particle production, accelerated expansion always ends for the temperature-dependent mechanism and potential $V_\text{I}$, as shown explicitly for $\gamma=5$ $H_0^{-1/4} M_\text{Pl}^{-3/4}$ in the right panel of Fig. \ref{fig:4}. Since $\Upsilon_\gamma = \gamma^2 T_\text{DR}^3$, the Hubble friction eventually overtakes the thermal friction when the dark radiation temperature decreases sufficiently. Once the thermal friction is negligible, the scalar field then approaches its simple quintessence attractor solution. Still, we find that a larger $\gamma$ causes $w_\text{TOT}$ to be smaller and less than $-\frac{1}{3}$ for a longer period, increasing the amount (number of $e$-folds), duration (time), and average magnitude $\left(\frac{\ddot a}{a}\right)$ of accelerated expansion. Despite there always being an end to accelerated expansion for this potential, the expansion rate is nearly indistinguishable from that of a cosmological constant up to present day for large enough $\gamma$.

Another consequence of the particle production's explicit temperature dependence is the current dark radiation density. Compared to the temperature-independent mechanism, the temperature-dependent mechanism can lead to a larger dark radiation temperature and density at present day, as illustrated in the right panel of Fig. \ref{fig:5}. As with $\beta$, a larger $\gamma$ does not necessarily correspond to a larger present-day dark radiation density. However, the temperature-dependent production rate $\Upsilon_\gamma\dot\varphi^2$ can be much larger on average than the temperature-independent production rate $\Upsilon_\beta\dot\varphi^2$. When compared to $\Upsilon_\beta$, the thermal friction coefficient $\Upsilon_\gamma$ can be smaller until closer to present day, resulting in less thermal friction and allowing $\dot\varphi$ to be larger. At this point, $\Upsilon_\gamma$ can grow rapidly, thereby leading to a larger on-average production rate $\Upsilon_\gamma\dot\varphi^2$. As a result, a larger present-day dark radiation density is theoretically possible with the temperature-dependent mechanism.
\section{End of Cosmic Expansion $(V=V_\textmd{II})$}
\label{sec:end}
\allowdisplaybreaks[4]

\begin{figure}[t]
    \centering
    \includegraphics[width=0.8\textwidth]{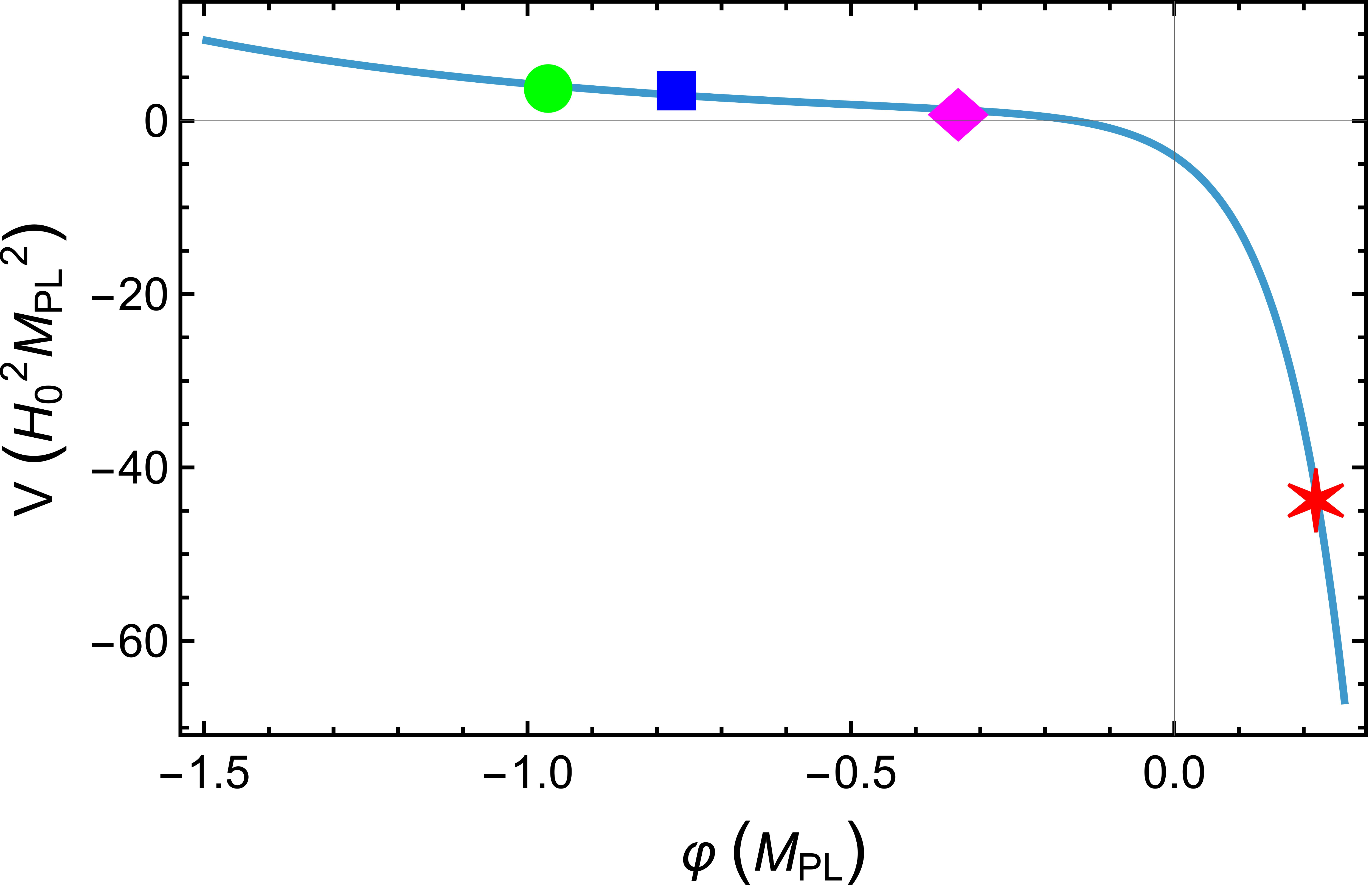}
    \caption{Difference of exponentials quintessence potential $V_\text{II}(\varphi)$ with $V_1 = 1$ $H_0^2M_{\text{Pl}}^2$, $V_2 = 5$ $H_0^2M_{\text{Pl}}^2$, $\lambda_1 = 1.5$, and $\lambda_2 = 10$. The green dot and blue square show where numerical integration and accelerated expansion begin, respectively, and the magenta diamond and red star show where accelerated expansion and cosmic expansion end, respectively, assuming no particle production.}
    \label{fig:6}
\end{figure}

We next consider $V_\text{II}(\varphi)=V_1 e^{-\lambda_1\varphi}-V_2e^{\lambda_2\varphi}$ with $\lambda_1\geq \sqrt{2}$ and $\lambda_2>\sqrt{6}$. Unlike $V_\text{I}$, $V_\text{II}$ ranges from positive to negative values, as shown in Fig. \ref{fig:6}. As the scalar field evolves down this potential, the universe undergoes a series of transitions from accelerated expansion to decelerated expansion to the end of expansion and beginning of contraction.

As before, we numerically solve the cosmological equations of motion, Eqs. (\ref{eq:1})$-$(\ref{eq:4}), for each particle production mechanism. For potential $V_\text{II}$, however, we stop integration at the end of cosmic expansion, when $H=0$. Starting from the final conditions of the numerical solutions during expansion at $H=0$, we then flip the sign of the Hubble parameter and numerically solve the equations of motion as the universe contracts. In this paper, we focus predominantly on the evolution of the universe during the expansion phase. For the plots in this section, we set $V_1 = 1$ $H_0^2M_{\text{Pl}}^2$, $V_2 = 5$ $H_0^2M_{\text{Pl}}^2$, $\lambda_1 = 1.5$, and $\lambda_2 = 10$.

\subsection{Temperature-Independent Particle Production Mechanism $\left(\Upsilon=\Upsilon_\beta\right)$}

As with potential $V_\text{I}$, we find that thermal friction from particle production reduces $\ddot\varphi$. As a result, $w_\varphi$ increases more slowly, and the average magnitude of accelerated expansion $\frac{\ddot a}{a}$ is greater.
\begin{figure}[t]
    \centering
    \begin{minipage}{.5\textwidth}
      \centering
     \includegraphics[width=1\linewidth]   {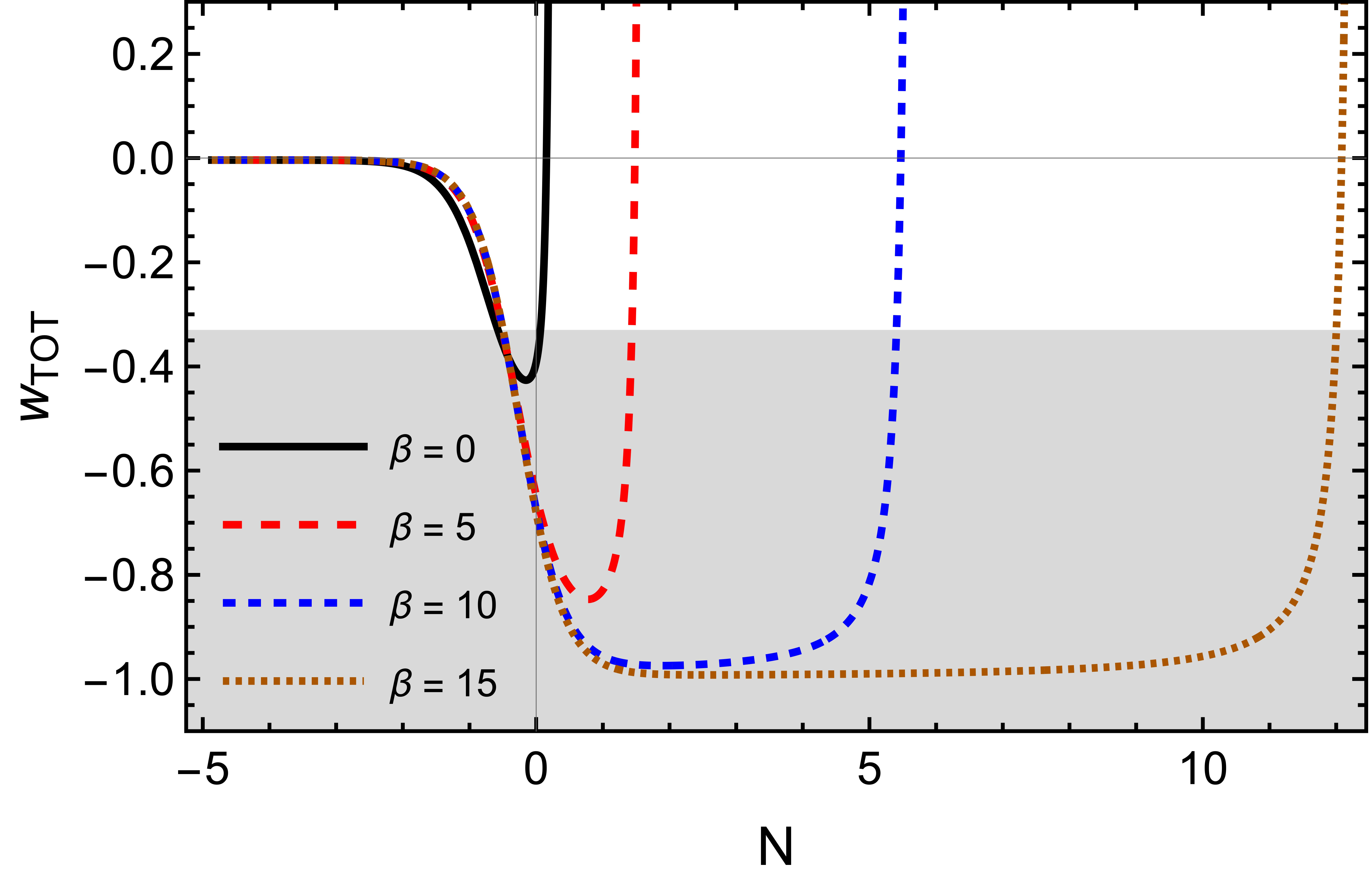}
    \end{minipage}%
    \begin{minipage}{.5\textwidth}
      \centering
      \includegraphics[width=1\linewidth]{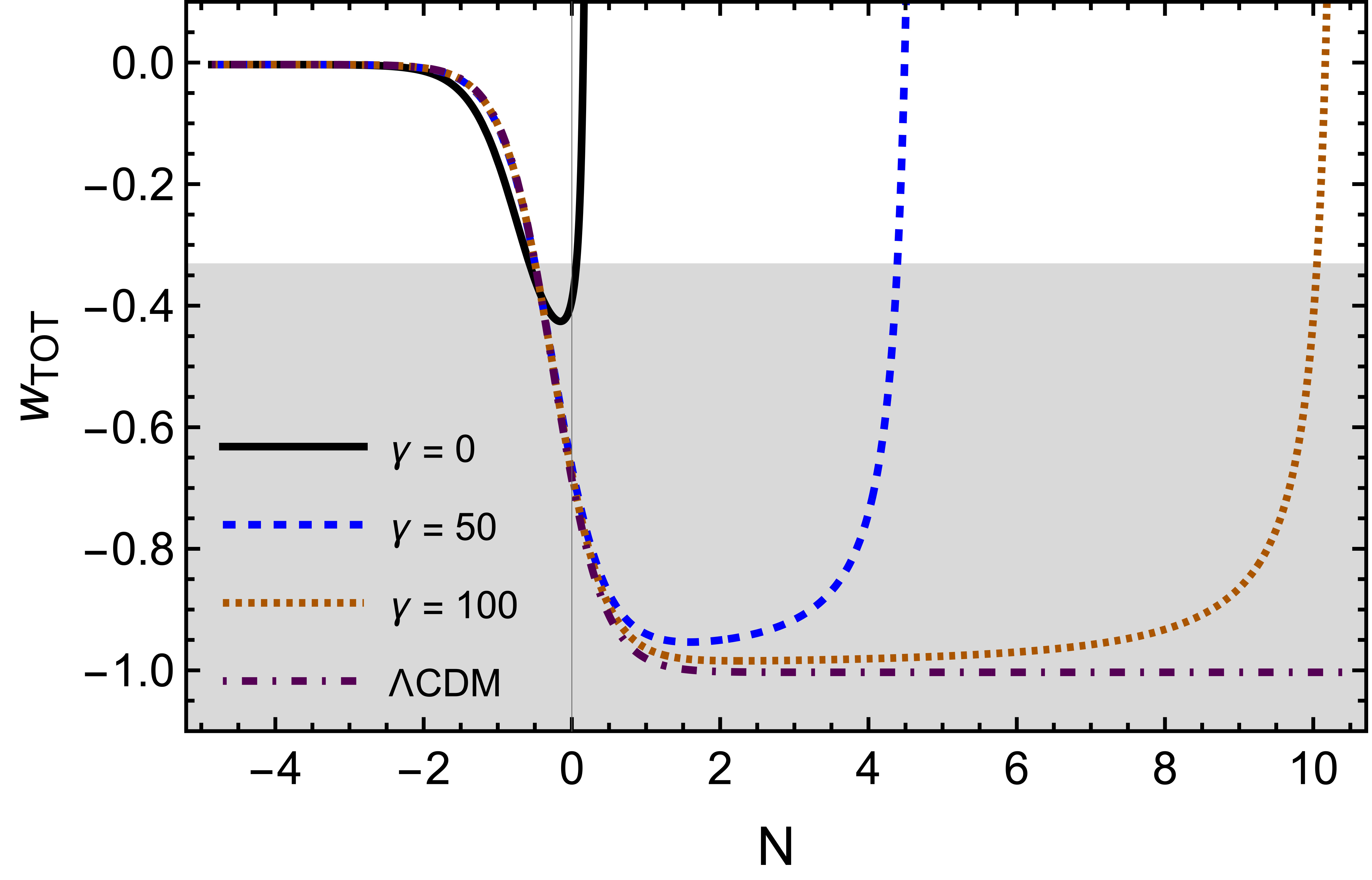}
    \end{minipage}
    \caption{Total equation of state $w_\text{TOT}$ as a function of $N$ $e$-folds relative to present day $(N=0)$ with difference of exponentials potential $V_\text{II}$ for the temperature-independent (left panel) and temperature-dependent (right panel) particle production mechanisms. Both plots compare the case of simple quintessence ($\beta=\gamma=0$) against cases where $\beta$ (unitless) and $\gamma$ (in units of $H_0^{-1/4}M_{\text{Pl}}^{-3/4}$) are non-zero, and the right panel includes a curve for $\Lambda$CDM. When a curve is in the shaded region $\left(w_\text{TOT} < -\frac{1}{3}\right)$, the universe is undergoing accelerated expansion. Cosmic expansion ends for each case when $w_\text{TOT}$ diverges, so larger $\beta$ and $\gamma$ delay the end of cosmic expansion.}
    \label{fig:7}
\end{figure}
For large enough $\beta$, steep TCC-allowed potentials, such as $V_\text{II}$, can have accelerated expansion similar to what occurs for flatter potentials or a cosmological constant.

Unlike for $V_\text{I}$, temperature-independent particle production does not lead to eternal accelerated expansion. The scalar field always reaches a sufficiently negative region of the potential for accelerated and decelerated cosmic expansion to end. As shown in the left panel of Fig. \ref{fig:7}, we find that a larger $\beta$ results in more $e$-folds of accelerated expansion before expansion ends, where $w_\text{TOT}$ diverges. This increased amount and duration of accelerated expansion consequently increases the time until contraction begins. \textit{Thus, thermal friction from particle production delays the end of cosmic expansion.}

Many of the results for the temperature-independent mechanism and potential $V_\text{I}$ are still true for $V_\text{II}$. The present-day dark radiation density again does not increase monotonically with $\beta$. A slightly larger $\Omega_\text{DR,0}$ is possible for $V_\text{II}$ compared to $V_\text{I}$, given the same $V_1$ and $\lambda_1$, but this difference is expected since $V_\text{II}$ is steeper than $V_\text{I}$. Further, one can obtain the same $\Omega_\text{DR,0}$ for different combinations of the shape of the potential, the steepness of the potential, and the particle production mechanism, making it challenging to differentiate a steeper single exponential from a flatter difference of exponentials potential. However, in combination with additional cosmological data, measurements of $\Omega_\text{DR,0}$ may indicate the shape of the potential and whether cosmic expansion will end.

\subsection{Temperature-Dependent Particle Production Mechanism $\left(\Upsilon=\Upsilon_\gamma\right)$}

For temperature-dependent particle production, many of the previously discussed results still hold for $V_\text{II}$. As with $V_\text{I}$, $w_\varphi(z)$ can have a sizable bump when the field is sufficiently decelerated by the combined thermal and Hubble friction, similar to what is shown in the right panel of Fig. \ref{fig:2}. Thermal friction still increases the amount, duration, and average magnitude $\left(\frac{\ddot a}{a}\right)$ of accelerated expansion and allows steep potentials, such as $V_\text{II}$, to produce similar accelerated expansion to a cosmological constant. We see in the right panel of Fig. \ref{fig:7} that a larger $\gamma$ further delays the end of cosmic expansion, and the total equation of state $w_\text{TOT}$ temporarily approaches the $\Lambda$CDM $w_\text{TOT}$ for large enough $\gamma$. We again find that $V_\text{II}$ and temperature-dependent thermal friction allow for a higher possible current dark radiation density, given the same $V_1$ and $\lambda_1$.

\begin{figure}[t]
    \centering
    \begin{minipage}{.5\textwidth}
      \centering
     \includegraphics[width=1\linewidth]   {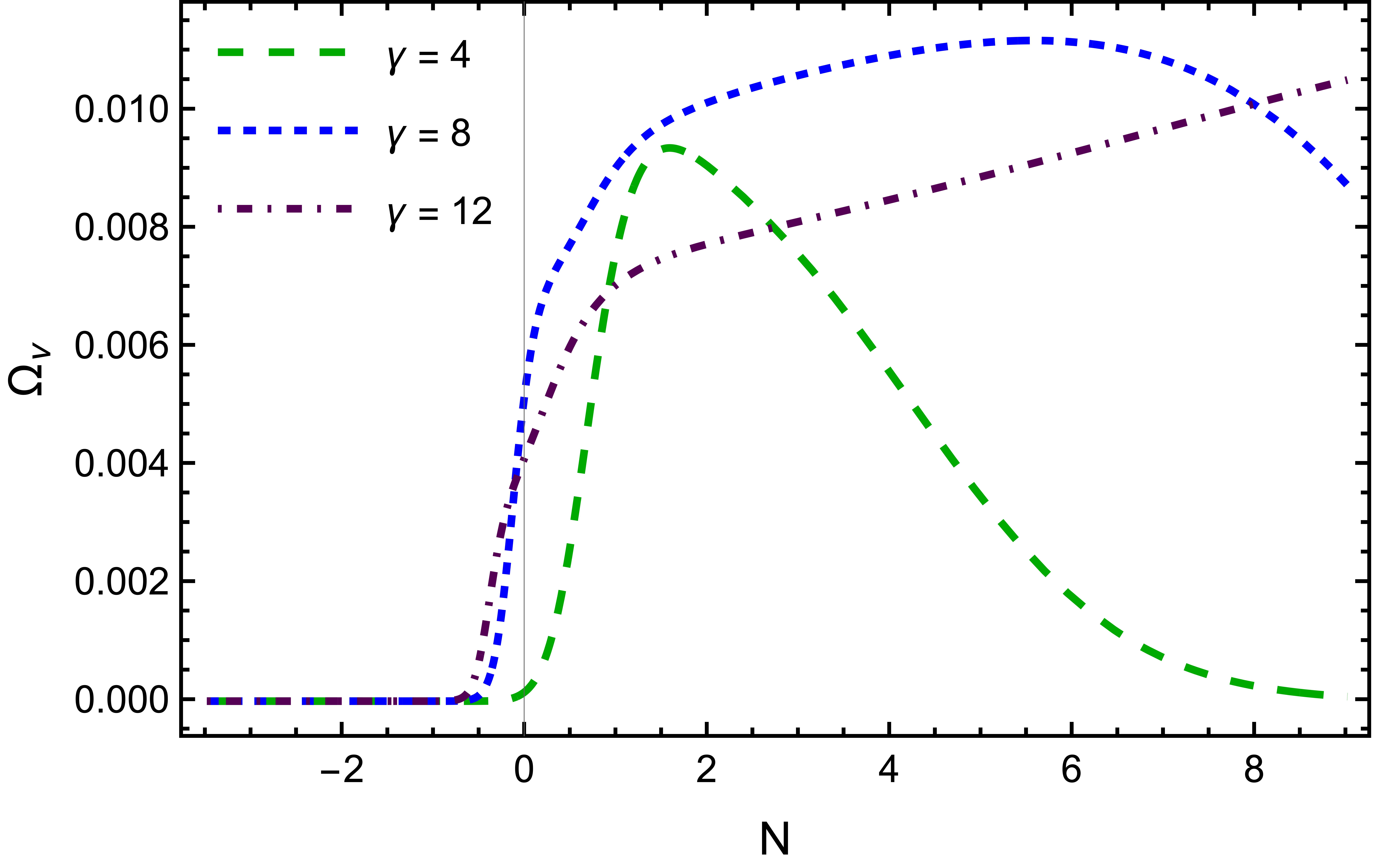}
    \end{minipage}%
    \begin{minipage}{.5\textwidth}
      \centering
      \includegraphics[width=1\linewidth]{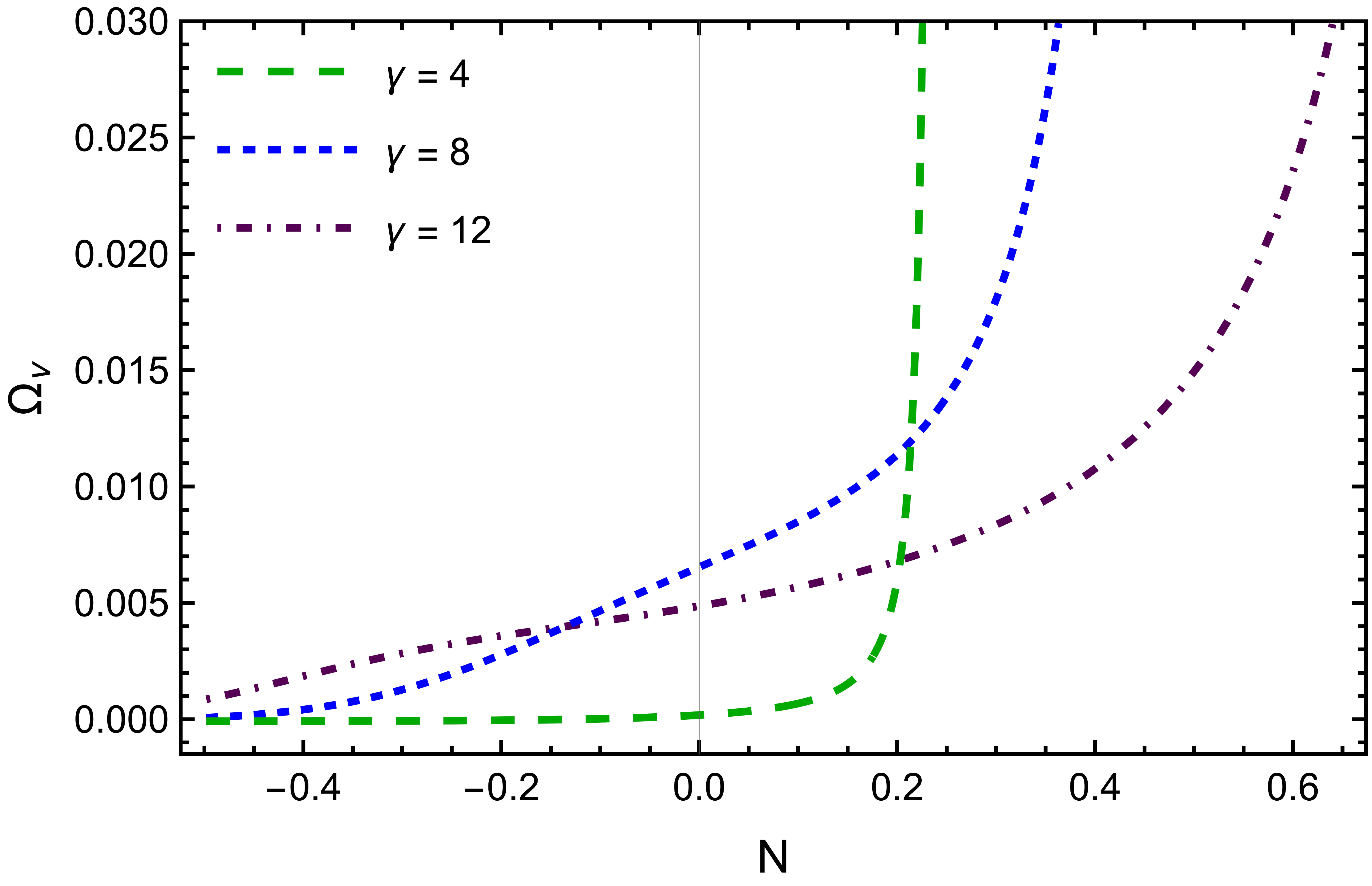}
    \end{minipage}
    \caption{Lightest Standard Model neutrino fractional density $\Omega_\nu= \frac{1.75}{21}\frac{\rho_{\text{DR}}}{3H^2}$ as a function of $N$ $e$-folds relative to present day $(N=0)$ for a steep exponential potential $V_\text{I}$ (left panel) and difference of exponentials potential $V_\text{II}$ (right panel) up to the end of cosmic expansion for each $\gamma$ (in units of $H_0^{-1/4}M_{\text{Pl}}^{-3/4}$). $\Omega_\nu$ diverges when cosmic expansion ends. For the cases considered, $\Omega_\nu$ reaches a maximum value in the future for $V_\text{I}$, where cosmic expansion is eternal, but increases until the end of cosmic expansion for $V_\text{II}$.}
    \label{fig:8}
\end{figure}

As discussed in Section \ref{sec:3}, the lightest Standard Model neutrino could be part of the dark radiation background. Since this neutrino is a fermion with two polarizations, the fractional neutrino density is simply $\Omega_\nu = \frac{1.75}{21}\Omega_\text{DR}$, assuming $g_* = 21$. The evolution of $\Omega_\nu$ up to present day and beyond is sensitive to the shape of the potential, as illustrated in Fig. \ref{fig:8}.
For a steep exponential potential $V_\text{I}$, $\Omega_\nu$ reaches a maximum value and then decreases as the scalar field approaches its simple quintessence attractor solution. For a difference of exponential potentials $V_\text{II}$, past present day, $\Omega_\nu$ increases until the end of cosmic expansion. Though $\Omega_\nu$ diverges at this point, the physical neutrino density $\rho_\nu$ always remains finite. Through experiments such as PTOLEMY \cite{ptolemy}, measurements of this neutrino density may indicate whether the end of cosmic expansion is nigh.

During contraction, we find that, for large enough $\gamma$, the dark radiation density can surpass the scalar field energy density and dominate the universe for as long as temperature-dependent particle production holds. As the universe contracts, the temperature of the dark radiation increases rapidly. In combination with an explicitly temperature-dependent production rate $\gamma^2 T_\text{DR}^3\dot\varphi^2$, the dark radiation density can grow large enough to prevent the scalar field from ever taking over. 

To show this analytically, we consider a contracting universe dominated by dark radiation. We approximate the Hubble parameter as $H\approx -\sqrt{\frac{1}{3}\rho_\text{DR}}=-T_\text{DR}^2\sqrt{\frac{g_*\pi^2}{90}}$. Then, the dark radiation density grows as
\begin{equation}
    \begin{split}
        \dot\rho_\text{DR}&= -4H\rho_\text{DR}+\gamma^2 T_\text{DR}^3\dot\varphi^2\\
        &\approx \frac{4}{\sqrt{3}}\left(\frac{g_*\pi^2}{30}\right)^{3/2} T_\text{DR}^6+\frac{g_*\pi^2\gamma^2}{90}T_\text{DR}^7\varphi'^2,
    \end{split}
\end{equation}
where we use that $\dot\varphi=H\varphi'$, and the scalar field energy density grows as
\begin{equation}
    \begin{split}
        \dot\rho_\varphi &=-3H\dot\varphi^2-\gamma^2T_\text{DR}^3\dot\varphi^2\\
        &\approx \frac{1}{\sqrt{3}}\left(\frac{g_*\pi^2}{30}\right)^{3/2}T_\text{DR}^6\varphi'^2-\frac{g_*\pi^2\gamma^2}{90}T_\text{DR}^7\varphi'^2.
    \end{split}
\end{equation}
Then, $\dot\rho_\text{DR}>\dot\rho_{\varphi}$ when
\begin{equation}
    T_\text{DR}>\frac{1}{2\gamma^2}\sqrt{\frac{g_*\pi^2}{10}}\left(1-\frac{4}{\varphi'^2}\right).
\end{equation}
Once $T_\text{DR}>\frac{1}{2\gamma^2}\sqrt{\frac{g_*\pi^2}{10}}$, the dark radiation density will always grow faster than the scalar field energy density, regardless of $\varphi'$ or the steepness of the potential, because the dark radiation temperature always increases in a contracting universe. Thus, for large enough temperatures in a contracting, dark radiation-dominated universe, the scalar field will never come to dominate. For $\gamma \sim H_0^{-1/4} M_\text{Pl}^{-3/4}$, the temperature must be $T_\text{DR}\gtrsim H_0^{1/2}M_\text{Pl}^{1/2}$ for this to occur, which we see is possible in our numerical solutions. 

In some models of cyclic cosmology \cite{cyclic}, slow contraction via scalar field, where $w_\varphi>1$, is responsible for smoothing and flattening the universe before the next bounce. This process requires that the scalar field eventually dominates during contraction, which may not occur if there is temperature-dependent particle production. However, simply because the scalar field does not ever dominate does not necessarily mean that slow contraction is prevented. If dark radiation grows with an effective equation of state $w_\text{DR, eff}>1$, then slow contraction can occur via the dark radiation instead.

For ordinary radiation, the continuity equation can be written as 
\begin{equation}
    \dot\rho_r=-3H\rho_r(1+w_r),
\end{equation}
where $w_r=\frac{1}{3}$. We can find the effective equation of state for the dark radiation, which accounts for particle production, by rewriting its continuity equation in this form:
\begin{equation}
    \dot\rho_\text{DR}= -3H\rho_\text{DR}\left(1+\frac{1}{3}-\frac{10\gamma^2H\varphi'^2}{g_*\pi^2T_\text{DR}}\right),
\end{equation}
so
\begin{equation}
    w_\text{DR, eff} = \frac{1}{3}-\frac{10\gamma^2H\varphi'^2}{g_*\pi^2T_\text{DR}}.
\end{equation}
Slow contraction via dark radiation occurs when dark radiation dominates and $w_\text{DR, eff}>1$, so in a contracting universe dominated by dark radiation, the temperature must be
\begin{equation}
    T_\text{DR}>\frac{2}{\gamma^2\varphi'^2}\sqrt{\frac{g_*\pi^2}{10}}
\end{equation}
for slow contraction to occur. 

In the cases we considered, the dark radiation temperature is never sufficiently large for dark radiation-driven slow contraction to occur. When the temperature is large enough to prevent the scalar field from dominating, we find that $\varphi'\rightarrow0$ and $w_\text{DR, eff}\rightarrow\frac{1}{3}$, so as long as temperature-dependent particle production occurs in these cases, contraction is not slow. However, whether this behavior will continue eternally for increasingly large temperatures is uncertain. As contraction continues, the temperature of the dark radiation will continue to increase rapidly. The temperature-dependent particle production mechanism may break down at large enough temperatures, such as those greater than the symmetry breaking scale $f$. Ultimately, the future of the contracting universe depends on the asymptotic shape of the potential and other components of the theory.
\section{Discussion}
\label{sec:discussion}
\allowdisplaybreaks[4]

This paper has examined various possible consequences of time-varying dark energy due to a quintessence scalar field whose energy density is partially converted to particles that comprise a background of thermal dark radiation. We considered two particle production mechanisms: a generic temperature-independent mechanism $(\Upsilon=\Upsilon_\beta)$, where $\Upsilon_\beta$ is a constant, and a temperature-dependent mechanism $(\Upsilon=\Upsilon_\gamma)$, where $\Upsilon_\gamma$ depends explicitly on the dark radiation temperature \cite{DER,cosmology_der}. These particle production mechanisms result in thermal friction on the scalar field that can make it difficult to distinguish between a radiating field with a steep potential, a self-interacting field with a flatter potential, or a cosmological constant. We numerically solved the cosmological equations of motion for each particle production mechanism and for two types of potentials: a steep exponential potential $V_\text{I}$, where cosmic expansion is eternal, and a difference of exponentials potential $V_\text{II}$ that ranges from positive to negative values, where cosmic expansion ends. We summarize the results below:
\begin{itemize}
    \item For all potentials and particle production mechanisms, the amount of $e$-folds, duration, and average magnitude $\left(\frac{\ddot a}{a}\right)$ of accelerated expansion is increased by thermal friction. For sufficiently large thermal friction, steep TCC-allowed potentials, such as $V_\text{I}$ and $V_\text{II}$, can have accelerated expansion similar to what occurs for flatter potentials or a cosmological constant.
    \item For either potential and $\Upsilon=\Upsilon_\gamma$, $w_\varphi(z)$ can have a sizable bump before present day.
    \item For $V=V_\text{I}$ and $\Upsilon=\Upsilon_\beta$, present accelerated expansion either continues eternally or transitions to a temporary phase of decelerated expansion, depending on the value of $\Upsilon_\beta$. In cases where present accelerated expansion ends, an eternal era of accelerated expansion begins once thermal friction becomes significant relative to Hubble friction.
    \item For $V=V_\text{I}$ and $\Upsilon=\Upsilon_\gamma$, present accelerated expansion ends, but cosmic expansion is eternal.
    \item For $V=V_\text{II}$ and either particle production mechanism, the end of cosmic expansion is delayed by thermal friction, but cosmic expansion always transitions to contraction. 
    \item For $V=V_\text{II}$ and $\Upsilon=\Upsilon_\gamma$, slow contraction can be prevented by dark radiation domination.
\end{itemize}

For a purely self-interacting scalar field, the Trans-Planckian Censorship Conjecture (TCC) limits not only the allowed flatness of the scalar potential $\left(\frac{\left|V'\right|}{V}\geq \sqrt{2}\right)$ but also the amount of $e$-folds of accelerated expansion \cite{TCC}. Though particle production can cause steep TCC-allowed potentials to be in better agreement with dark energy observational constraints, the increased amount of accelerated expansion may still violate the TCC. If there is a sufficiently large background of radiation through particle production, such as with warm inflation or a radiating quintessence field, whether the TCC limits the amount of accelerated expansion is uncertain. However, in the limit where the radiation density vanishes relative to the scalar field energy density, TCC constraints on the amount of accelerated expansion and steepness of the potential still apply. For $V=V_\text{I}$ and $\Upsilon=\Upsilon_\beta$, where accelerated expansion is eternal, we find that the scalar field energy density dominates $(\Omega_\varphi\rightarrow1)$ in the asymptotic future. Thus, a steep exponential potential with temperature-independent particle production is forbidden by the TCC.

Finally, the production of a background of dark radiation provides interesting detection prospects for time-varying dark energy. For all potentials and particle production mechanisms, we find that the present dark radiation density can be much larger than the remnant photon energy density. If the lightest Standard Model neutrino is part of this dark radiation, there could be a neutrino background at temperatures higher than the relic C$\nu$B which may be detectable by experiments such as PTOLEMY \cite{ptolemy}, as explored in \cite{cosmology_der}. We showed that evolution of the dark radiation fractional density, and thus the neutrino fractional density, differs depending on the shape of the potential. We leave to future work to determine whether dark radiation measurements in combination with current cosmological data can distinguish between quintessence potentials and ascertain whether the end of cosmic expansion is nigh.
\section*{Acknowledgements} \label{sec:acknowledgements}
We thank Anirudh Prabhu, Kim Berghaus, and Larry Yaffe for many useful conversations. This work is supported in part by the DOE grant number DEFG02-91ER40671 and by the Simons Foundation grant number 654561.
\bibliographystyle{references/JHEP.bst}
\bibliography{references/references}

\end{document}